# Comparison of the earliest NC and CC planetesimals: Evidence from ungrouped iron meteorites

Fridolin Spitzer[1,2], Christoph Burkhardt[1,2], Thomas S. Kruijer[3], and Thorsten Kleine[1,2]

[1]Max Planck Institute for Solar System Research, Department for Planetary Sciences, Justus-von-Liebig-Weg 3, 37077 Göttingen, Germany.

[2]Institut für Planetologie, University of Münster, Wilhelm-Klemm-Str. 10, 48149 Münster, Germany.

[3]Nuclear & Chemical Sciences Division, Lawrence Livermore National Laboratory, 7000 East Avenue (L-231), Livermore, CA 94550, USA.

*Corresponding author: Fridolin Spitzer (spitzer@mps.mpg.de)

# Abstract

Isotope anomalies in meteorites reveal a fundamental dichotomy between *non-carbonaceous-* (NC) and *carbonaceous-type* (CC) planetary bodies. Until now, this dichotomy is established for the major meteorite groups, representing about 36 distinct parent bodies. Ungrouped meteorites represent an even larger number of additional parent bodies, but whether they conform to the overall NC-CC dichotomy is unknown. Here, the genetics and chronology of 26 ungrouped iron meteorites is considered through nucleosynthetic Mo and radiogenic W isotopic compositions. Secondary cosmic ray-induced modifications of these isotope compositions are corrected using Pt isotope measurements on the same samples. We find that all of the ungrouped irons have Mo isotope anomalies within the range of the major meteorite groups and confirm the NC-CC dichotomy for Mo, where NC and CC meteorites define two distinct, subparallel *s*-process mixing lines. All ungrouped NC irons fall on the NC-line, which is now precisely defined for 41 distinct parent bodies. The ungrouped CC irons show scatter around the CC-line indicative of small *r*-process Mo heterogeneities among these samples. These *r*-process Mo isotope variations correlate with O isotope anomalies, most likely reflecting mixing of CI chondrite-like matrix, chondrule precursors and Ca-Al-rich inclusions. This implies that CC iron meteorite parent bodies accreted the same nebular components as the later-formed carbonaceous chondrites. The Hf-W model ages of core formation for the ungrouped irons overlap with those of the iron meteorite groups from each reservoir and reveal a narrow age peak at ~3.3 Ma after Ca-Al-rich inclusions for the CC irons. By contrast, the NC irons display more variable ages, including younger ages indicative of impact-induced melting events, which seem absent among the CC irons. This is attributed to the more fragile and porous nature of the CC bodies, making impact-induced melting on their surfaces difficult. The chemical characteristics of all iron meteorites together reveal slightly more oxidizing conditions during core formation for CC compared to NC irons. More strikingly, strong depletions in moderately volatile elements, typical of many iron meteorite parent bodies, predominantly occur among CC irons, for reasons that remain unclear at present.

# 1 Introduction

Nucleosynthetic isotope anomalies in bulk meteorites reflect the heterogeneous distribution of isotopically diverse presolar materials in the early solar nebula and have been identified for a large number of elements (see summaries in e.g. Kruijer et al., 2020; Kleine et al., 2020; Bermingham et al., 2020; Burkhardt et al., 2021; Dauphas et al., 2024). In general, these anomalies reveal a fundamental isotopic dichotomy between *non-carbonaceous* (NC) and *carbonaceous* (CC) materials, which are presumed to represent the inner and outer solar system, respectively (Warren, 2011; Budde et al., 2016a). Of the elements defining the NC-CC dichotomy, Mo is particularly important for two reasons. First, for Mo the dichotomy is very clearly resolved, because when the anomalies in $^{94}$Mo and $^{95}$Mo are plotted against each other, meteorites fall along two approximately parallel lines with a nearly constant offset between the two lines (Budde et al., 2016a; Kruijer et al., 2017; Bermingham et al., 2018; Budde et al., 2019; Yokoyama et al., 2019; Spitzer et al., 2020). The isotopic variations along each of these two lines (termed the NC- and CC-lines; Budde et al., 2016a) predominantly reflect variation in *s*-process Mo, while the offset between the lines is due to an excess in *r*-process Mo in the CC over the NC reservoir. Second, Mo isotopes can be readily measured in iron meteorites. This is important because the accretion time of iron meteorite parent bodies can be determined through $^{182}$Hf-$^{182}$W dating of core formation and thermal modeling, and is <1 Ma after formation of Ca-Al-rich inclusions (CAIs) for most NC and CC irons (Kruijer et al., 2014b, 2017; Hilton et al., 2019; Tornabene et al., 2020; Spitzer et al., 2021; Tornabene et al., 2023; Chiappe et al., 2023b). As such, the iron meteorite data demonstrate that the NC and CC nebular reservoirs have been established and separated from each other by ~1 Ma after CAI formation at the latest (Budde et al., 2016a; Kruijer et al., 2017).

Until now, the NC-CC dichotomy has been studied almost exclusively using samples from the well-defined meteorite groups. These represent ~16 undifferentiated and ~24 differentiated parent bodies (Krot et al., 2013). This number is small, however, when compared to the 395 meteorites currently listed as ungrouped in the Meteoritical Bulletin Database (98 chondrites, 150 iron meteorites, 147 achondrites), which—even after taking potential pairings into account—sample a much larger number of parent bodies than the meteorite groups (Goldstein et al., 2009; Greenwood et al., 2020). This means that so far only less than a quarter of the planetary bodies sampled by meteorites have been investigated for their nucleosynthetic heritage. Given the considerable potential of this untapped sample reservoir for better understanding the nucleosynthetic isotope heterogeneity of the early solar system, we initiated a systematic isotopic and chronological study of ungrouped iron meteorites.

Iron meteorites can either be of "magmatic" or "non-magmatic" origin. The former are thought to sample the fractional crystallization sequence of metallic cores of differentiated objects (Scott, 1972), where each chemical group represents a distinct parent body. The origin and number of parent bodies represented by the "non-magmatic" or "silicate-bearing groups" (IAB complex, IIE group) are less certain because their chemical compositions are not easily accounted for by fractional crystallization alone (Scott, 1972; Goldstein et al., 2009). Thus, alternative formation mechanism(s) appear to be required for these groups, such as formation in large, impact-induced melt pools (e.g., Wasson and Wang, 1986). Ungrouped iron meteorites comprise a diverse set of samples, including both magmatic and non-magmatic samples, whose chemical compositions cannot be matched to any of the established groups via fractional crystallization models. To first order, each ungrouped iron meteorite is thought to represent a distinct parent body, which makes them ideal for investigating a large set of different parent bodies.

We report Mo and W isotope data for 26 ungrouped iron meteorites. Since the original isotopic composition of meteorites can be severely modified by secondary neutron capture during cosmic ray exposure (CRE), Pt isotope compositons were also measured for every sample and are used to correct the measured W and Mo isotopic compositions for CRE effects (Kruijer et al., 2013; Spitzer et al., 2020). The new Mo and W isotope data are used to test whether the NC-CC dichotomy extends to the large number of parent bodies sampled by these meteorites and to better understand any systematic differences between the NC and CC reservoirs, and the parent bodies formed within each reservoir.

# 2 Samples and analytical methods

### 2.1 Iron meteorite samples

Twenty-six ungrouped iron meteorites were selected for this study (Table 1). They were obtained from the Field Museum Chicago, the Natural History Museum London, the Smithsonian Institution, the Max Planck Institute (MPI) for Nuclear Physics in Heidelberg, and the Institut für Planetologie in Münster. To assess as to whether chemical variations among iron meteorites are related to their genetics (i.e., whether they are NC or CC), a chemically diverse sample set was selected (*e.g.*, Fig. 1 and Table S1). For instance, Tishomingo has one of the highest Ni concentrations (~32 wt%) among iron meteorites, while the Ge contents of NWA 859 and N'Goureyma differ by five orders of magnitude. In addition, samples for which O isotopic data exist were chosen preferentially to examine if they correlate with the Mo isotopic compositions (Table 1).

We also attempted to include both magmatic and non-magmatic irons within the sample set of this study as these two groups of irons presumably formed by different processes. All known non-magmatic irons for which genetic information is available thus far are NC, raising the question of whether or not the non-magmatic irons are exclusive to the NC reservoir. To distinguish between a magmatic or non-magmatic origin of individual samples, we followed Wasson (2013) who postulated that an Au content ≥0.3 µg/g and an Ir/Au ratio far from chondritic values (≤ 1 or ≥ 9) are indicative of a magmatic origin. On this basis, 8 of the 26 ungrouped irons of this study may be of non-magmatic origin (Tables 1 and S1). In addition to these chemical and petrographic characteristics, the isotopic data of this study will help distinguish between a sample's magmatic or non-magmatic origin.

### 2.2 Chemical separation and isotope measurements

The iron meteorite samples were cut using a diamond saw, polished with abrasives (SiC) to remove saw marks and subsequently ultrasonically cleaned in ethanol to get rid of any adhering dust. Individual samples weighing ~0.4−0.6 g were digested in 6M HCl (+trace $HNO_3$) on a hot plate (130 °C) for at least 24 h. As neutron capture effects in a single iron meteorite can be strongly variable (Kruijer et al., 2013; Leya and Masarik, 2013), W, Mo, and Pt isotope analyses were performed on aliquots from a single digestion. Upon complete dissolution, an aliquot representing 50 mg of material was taken for Pt isotope analyses, while the remainder was processed for W and Mo isotope analyses. The chemical separation and isotope measurements of Pt followed the protocols described in Kruijer et al. (2013) and are based on techniques initially described by Rehkämper and Halliday (1997). Platinum is separated from the sample matrix using an anion exchange chromatography step, where total yields for Pt were typically ~80%. The separation of W and Mo from the remaining solution (~0.5 g) followed the analytical protocol as described in (Kruijer et al., 2017; Budde et al., 2018), which will be briefly summarized below (see Appendix for details). The separation of W for isotope composition analyses was achieved using a two-stage anion exchange chromatography, where the yield was typically ~60%. Molybdenum was collected during the two-stage anion exchange chemistry used for the separation of W and subsequently purified using a two-stage ion exchange chromatography slightly modified after (Burkhardt et al., 2011, 2014), where the Mo yield for the entire procedure was typically ~75%. The mean total procedural blanks for Mo (3.9 ng), W (0.05 ng), and Pt (0.18 ng) were negligible for all samples.

The Mo, W, and Pt isotope measurements were performed on the Thermo Scientific Neptune Plus MC-ICP-MS in the Institut für Planetologie at the University of Münster. Isobaric interferences of Zr and Ru on Mo masses were corrected by monitoring $^{91}Zr$ and $^{99}Ru$. Instrumental mass bias was corrected by internal normalization to $^{98}Mo/^{96}Mo$ = 1.453173, $^{186}W/^{184}W$ = 0.92767 (denoted '6/4') or $^{186}W/^{183}W$ = 1.98590 (denoted '6/3'), and to either $^{196}Pt/^{195}Pt$ = 0.7464 (denoted '6/5') or $^{198}Pt/^{195}Pt$ = 0.2145 (denoted '8/5') using the exponential law. All isotopic data are reported in the $\varepsilon$-notation, i.e., the parts per $10^4$ deviation relative to terrestrial bracketing Alfa Aesar solution standards. For samples analyzed several times, reported values represent the mean of pooled solution replicates. The accuracy and precision of the Mo, W, and Pt isotope measurements were assessed by repeated analyses of the NIST 129c and NIST 361 metal standards (Table S2; Spitzer et al., 2021), which had been doped with an Alfa Aesar standard solution containing highly siderophile elements (HSE) for Pt analyses. The measured isotopic compositions of the processed metal standards are indistinguishable from the solution standard analysis and consistent with previous studies (Kruijer et al., 2014b, 2017; Budde et al., 2018; Pape et al., 2023), testifying to the accuracy of our measurements.

Normalizations involving $^{183}W$ exhibit a small mass-independent effect, which has also been observed in several previous studies (*e.g.,* (Willbold et al., 2011; Kruijer et al., 2012) and results in small excesses for $\varepsilon^{182}W$ ('6/3'), $\varepsilon^{184}W$ ('6/3'), and corresponding deficits in $\varepsilon^{183}W$ ('6/4'). The magnitude of this analytical effect varies between different studies and is typically about -0.1 to -0.2 for $\varepsilon^{183}W$ and ascribed to a magnetic isotope effect induced during the chemical purification protocol (Budde et al., 2022). For all but three samples (see Appendix for details), W isotope ratios involving $^{183}W$ were therefore corrected using the mean $\varepsilon^{i}W$ values obtained for the metal standard analyzed in the present study, using the method described in (Kruijer et al., 2014a). The associated uncertainties induced through this correction were propagated and included in the reported uncertainties (see Appendix for details).

# 3 Results

## 3.1 Correction of CRE-effects on Mo and W isotopes

The CRE effects on Mo and W isotopes were monitored using Pt isotope measurement on digestion aliquots of the same samples. The Pt isotope data reveal variable CRE effects in the samples of this study, with $\varepsilon^{196}Pt$ varying from -0.07 to +0.49 (Table 2). Some samples display larger anomalies on $\varepsilon^{192}Pt$ (and $\varepsilon^{194}Pt$), reflecting the high Ir/Pt ratios of these samples, which lead to pronounced CRE-effects on $^{192}Pt$ and $^{194}Pt$ via *n*-capture on $^{191}Ir$ and $^{193}Ir$, respectively (Kruijer et al., 2013; Wittig et al., 2013). Only seven samples have elevated $\varepsilon^{196}Pt$ values (between 0.20 and 0.49), which require CRE corrections on their measured Mo and W isotope compositions. For the other samples these corrections are minor, but were made nevertheless.

The CRE-corrected, pre-exposure $\varepsilon^{i}Mo$ values for each sample were determined using the following equation:

$$\varepsilon^{i}Mo_{pre\text{-}exposure} = \varepsilon^{i}Mo_{measured} - [\varepsilon^{196}Pt_{measured} + 0.06 \pm 0.01] \times m(\varepsilon^{i}Mo \text{ vs. } \varepsilon^{196}Pt), \quad (1)$$

where i corresponds to 92, 94, 95, 97, or 100, respectively, $m(\varepsilon^{i}Mo$ vs. $\varepsilon^{196}Pt)$ is the updated empirically determined slope of $\varepsilon^{i}Mo$ vs. $\varepsilon^{196}Pt$ from this study (Table S3; see supplementary text for details), and the term +0.06±0.01 accounts for the correction of a small nucleosynthetic anomaly on $\varepsilon^{196}Pt$ (Spitzer et al., 2021).

For $^{182}W$, pre-exposure $\varepsilon^{182}W$ values were calculated similarly, using the following equation:

$$\varepsilon^{182}W_{pre\text{-}exposure} = \varepsilon^{182}W_{measured} - [\varepsilon^{196}Pt_{measured} + 0.06 \pm 0.01] \times m(\varepsilon^{182}W \text{ vs. } \varepsilon^{196}Pt), \quad (2)$$

where, as above, the term +0.06±0.01 accounts for the presence of small Pt nucleosynthetic isotope anomalies, and $m(\varepsilon^{182}W$ vs. $\varepsilon^{196}Pt)$ is the mean slope of the $\varepsilon^{182}W$ vs. $\varepsilon^{196}Pt$ correlation (-1.320 ± 0.055) obtained for the major iron meteorite groups (Kruijer et al., 2017). All associated errors have been propagated into the final uncertainties of the pre-exposure $\varepsilon^{i}Mo$ and $\varepsilon^{182}W$ values.

### 3.2 CRE-corrected Mo isotope results

The CRE-corrected Mo isotopic compositions of the 26 ungrouped iron meteorites are summarized in Table 3 and shown in Fig. 2. Five of these samples (Babb's Mill, ILD 83500, Grand Rapids, Mbosi, and Tishomingo) have been analyzed in prior studies, and the new data are in good agreement with these previous results (Dauphas et al., 2002; Burkhardt et al., 2011; Worsham et al., 2017; Hilton et al., 2019). All samples have positive $\varepsilon^{i}$Mo values, similar to Mo isotope anomalies previously reported for other meteorites and consistent with the observation that, compared to meteorites, the terrestrial Mo isotope composition is characterized by an *s*-process excess (e.g., Burkhardt et al., 2011). All samples of this study plot on either the NC- or the CC-line (Fig. 2), confirming the fundamental dichotomy between NC and CC materials defined by the major meteorite groups. The only exception thus far is the ungrouped iron meteorite Nedagolla, which has a mixed NC-CC Mo isotopic composition (Spitzer et al., 2022).

Sixteen of the 26 investigated ungrouped irons are associated with the CC reservoir, spanning a relatively restricted range of $\varepsilon^{94}$Mo values between ~1.15 and ~1.75. Tucson ($\varepsilon^{94}$Mo = 2.19 ± 0.08) is the only ungrouped CC iron with larger Mo isotope anomalies, which are similar to those of the IIC irons (Kruijer et al., 2017). Although all the ungrouped CC irons plot within uncertainty on the CC-line, several samples tend to plot slightly below this line (Fig. 2). The remaining 9 ungrouped irons of this study plot on the NC-line and have $\varepsilon^{94}$Mo varying from ~0.5 to ~1.3, which is slightly smaller than the range observed among the major meteorite groups.

### 3.3 CRE-corrected W isotope results

The W isotope data (after correction for the analytical $^{183}$W effect; see above and appendix for details) are summarized in Table 4 and shown in Fig. 3. For the ungrouped NC irons, the CRE-corrected $\varepsilon^{182}$W (6/4) values vary from -2.95 to -3.46, while the $\varepsilon^{183}$W (6/4) and $\varepsilon^{184}$W (6/3) are within uncertainty of zero. This and the consistency between the $\varepsilon^{182}$W (6/4) and $\varepsilon^{182}$W (6/3) values indicates that these NC irons have no resolved nucleosynthetic W isotope anomalies, consistent with the absence of such anomalies among NC meteorites (e.g., Kruijer et al., 2017). In contrast, the ungrouped CC irons show resolved $^{183}$W anomalies of nucleosynthetic origin, with $\varepsilon^{183}$W (6/4) values ranging from +0.09 to +0.26 (Fig. 3). This is in agreement with $^{183}$W excesses reported for the CC iron meteorite groups (Kruijer et al., 2017;

Hilton et al., 2019; Tornabene et al., 2020). Nucleosynthetic W isotope anomalies result in correlated $\varepsilon^{182}$W-$\varepsilon^{183}$W (6/4) and $\varepsilon^{182}$W-$\varepsilon^{184}$W (6/3) variations (Burkhardt et al., 2012a; Kruijer et al., 2014a; Budde et al., 2016b), and so the $\varepsilon^{182}$W of each sample can be corrected for nucleosynthetic variations using its $\varepsilon^{183}$W (or $\varepsilon^{184}$W). After quantifying and correcting for these nucleosynthetic and cosmogenic W isotope variations, the $\varepsilon^{182}$W (6/4) and $\varepsilon^{182}$W (6/3) values of each of the ungrouped CC irons agree, and define a restricted range of values around approximately -3.20 (Table 4), consistent with the $\varepsilon^{182}$W values reported for the CC iron meteorite groups (Fig. 3 and Table S4). For the ungrouped CC irons, the CRE-corrected $\varepsilon^{182}$W (6/4) values range from -3.02 to -3.29, while the $\varepsilon^{182}$W (6/3) vary from -3.05 to -3.33 (Table 4).

After correction for nucleosynthetic and CRE effects, the $\varepsilon^{182}$W values can be used to calculate a two-stage model age of metal segregation from an unfractionated, chondritic reservoir (e.g., Horan et al., 1998):

$$\Delta t_{CAI} = \frac{-1}{\lambda} ln\left[\frac{(\varepsilon^{182}W)_{sample} - (\varepsilon^{182}W)_{chondrites}}{(\varepsilon^{182}W)_{SSI} - (\varepsilon^{182}W)_{chondrites}}\right], \tag{3}$$

where $(\varepsilon^{182}W)_{sample}$ is the $\varepsilon^{182}$W of a sample, $(\varepsilon^{182}W)_{SSI}$ is the solar system initial value of -3.49 ± 0.07 obtained from CAIs (Kruijer et al., 2014a), $\lambda$ is the $^{182}$Hf decay constant of 0.0778 ± 0.0015 Ma$^{-1}$ (Vockenhuber et al., 2004), and $(\varepsilon^{182}W)_{chondrites}$ is the W isotopic composition of chondrites, assumed to represent the $\varepsilon^{182}$W value of the bulk, undifferentiated parent bodies. The mean $\varepsilon^{182}$W = -1.91 ± 0.08 (Kleine et al., 2004) of carbonaceous chondrites has traditionally been used for the latter, but recent work has shown that some ordinary (Hellmann et al., 2019) and the enstatite chondrites have distinctly lower $\varepsilon^{182}$W values (Hellmann et al., 2024). Of these, the bulk $\varepsilon^{182}$W = -2.23 ± 0.13 of enstatite chondrites is well-defined (Hellmann et al., 2024), and we will use this value to calculate model ages for the NC irons. For the CC irons we will use the mean $\varepsilon^{182}$W of carbonaceous chondrites instead. The Hf-W model ages calculated in this manner range from ~0.3 to ~7.2 Ma after CAI formation for the NC irons, and from ~1.5 to ~4.6 Ma for the CC irons (Fig. 4, Table 4).

# 4 Discussion

## 4.1 Implications for the NC–CC isotope dichotomy

The new Mo isotope data for ungrouped irons generally confirm the NC–CC dichotomy previously observed for the major iron meteorite groups. All ungrouped irons of this study fall within the range of nucleosynthetic Mo isotope anomalies known from the analyses of the iron meteorite groups, and they also plot on either the NC- or the CC-line in the $\varepsilon^{95}$Mo versus $\varepsilon^{94}$Mo diagram (Fig. 2). The only exception is Nedagolla, which has previously been shown to exhibit a mixed NC-CC Mo isotope signature, probably reflecting late-stage collisional mixing of genetically distinct materials (Spitzer et al., 2022). Thus, overall the Mo isotope data for ungrouped iron meteorites confirm the overarching NC-CC dichotomy of meteoritic materials for an additional 26 samples. This is important because most of these samples derive from distinct parent bodies, and so these data demonstrate that the NC-CC dichotomy extends to a large overall number of parent bodies (in total 41 distinct NC and 36 distinct CC bodies). Moreover, based on Hf-W chronometry it has been shown that the different groups of magmatic irons represent materials derived from bodies accreted within the first 1 Ma of the solar system. As the Hf-W model ages of the ungrouped irons of this study (with the expectation of a few NC samples of presumably non-magmatic origin; see Section 4.3) are similar to those of the magmatic groups, this also applies to the large number of parent bodies of the ungrouped irons. Together these data thus demonstrate that the NC-CC dichotomy was firmly established for many bodies within less than 1 Ma after solar system formation.

Although the ungrouped irons can be assigned to the NC or CC reservoirs, the new data reveal a systematic difference between the samples of either group. While the ungrouped NC irons all appear to plot precisely on the NC-line with no obvious scatter around the line, this does not seem to be the case for the CC irons, many of which appear to plot slightly below the CC-line. To assess this observation more quantitatively, we have updated the regression of the NC-line by including the new data for ungrouped irons. For this purpose, we compiled all available Mo isotope data for NC meteorites so that the NC regression is based on as many samples as possible. To avoid any systematic bias that may be introduced by utilizing different methods of CRE correction, we recalculated this correction for all irons for which the necessary data are available, using the same method as in this study (Table S5). The regression is calculated using one data point per parent body, so that a total of 41 distinct bodies are included in the regression (Fig. 5). Some uncertainty exists regarding the Mo isotopic composition of main-group IAB irons. Bermingham et al. (2025) recently argued that some of the Mo data obtained by prior N-TIMS may be inaccurate because of unaccounted-for fractionation effects. This issue is critical for the IAB irons, because most Mo isotope data for this group have been obtained using N-TIMS. In addition, the main-group IAB irons exhibit the smallest Mo isotope anomalies of all meteorites and as such define the lowermost point of the NC-line, exerting a strong control on the calculated slope of the NC-line. To circumvent the analytical issues that appear to be associated with some of the N-TIMS measurements on IAB irons, we use data from the IAB iron Campo del Cielo, which has been measured in several different studies and using both N-TIMS and MC-ICPMS (N=6; Poole et al., 2017; Worsham et al., 2017; Marti et al., 2023; Bermingham et al., 2025). The results for Campo de Cielo from all these studies are similar and given the different analytical techniques employed in these studies, it is quite unlikely that these data are compromised by analytical artefacts. Moreover, there are no significant CRE effects in Campo de Cielo (Worsham et al., 2017), making this sample well suited to determine the Mo isotope composition of IAB irons unaffected by CRE. In our data compilation, we, therefore, use the average Mo isotope composition of the individual analyses of Campo de Cielo reported in the aforementioned four studies to represent the Mo isotopic composition of the main-group IAB irons (Table S5).

We find that the data for the 41 distinct NC parent bodies define a single correlation line (MSWD = 1.3) with a slope of 0.517±0.028 (95% conf.) and intercept of -0.045±0.026 (95% conf.). These values are consistent with, albeit more precise (in most cases) than, previously reported values of 0.488±0.085 and -0.06±0.06 (Yokoyama et al., 2019), 0.528±0.045 and -0.058±0.045 (Spitzer et al., 2020), as well as 0.517±0.042 and 0.01±0.02 (Bermingham et al., 2025), all of which were based on a smaller sample set than in this study. Of note Bermingham et al. (2025) is the only study that reports a positive intercept for the NC-line, but this is because for the IAB irons these authors exclusively used data from their own study, and did not also include any of the previously published data for samples of this iron group. The slope of the NC-line determined here is slightly shallower than the slope of 0.596±0.008 obtained by Budde et al. (2019). This is because these authors also included Mo isotope data for acid leachates of ordinary chondrites in the regression. These leachates diplay much larger isotope anomalies than the bulk rocks and which predominantly reflect *s*-process variations. Thus, by including the leachate data in the regression, the slope naturally is shifted towards the slope expected for pure *s*-process variations, which is 0.596 (Budde et al., 2019). By contrast, bulk meteorites define a slightly shallower slope, indicating that while the variations among NC meteorites predominantly reflect *s*-process variability, this is coupled with minor variations in *r*-process nuclides (Spitzer et al., 2020).

In contrast to the well-correlated Mo isotope variations among the NC meteorites, the new data for the ungrouped CC irons exhibit more scatter (Fig. 6). One way to quantify this scatter is to use the $\Delta^{95}$Mo notation, which is the ppm-deviation of a sample's $^{94}$Mo and $^{95}$Mo isotope composition from an *s*-process mixing line passing through the origin and is calculated according to:

$$\Delta^{95}\text{Mo} = (\varepsilon^{95}\text{Mo} - 0.596 \times \varepsilon^{94}\text{Mo}) \times 100, \quad (4)$$

where 0.596 is the slope expected for pure *s*-process variations, which is determined based on Mo isotope data for acid leachates and presolar SiC grains (Fig. 2; Budde et al., 2019). Thus, variations in $\Delta^{95}$Mo are indicative of differences in the abundance of *r*-process Mo nuclides. The characteristic $\Delta^{95}$Mo value of CC meteorites has been determined to be $\Delta^{95}$Mo = 26±2 (Budde et al., 2019), but as shown in Fig. 6, for several ungrouped CC irons the $\Delta^{95}$Mo values extend towards lower values down to ~16, indicating *r*-process variations among the CC meteorites. These variations are small compared to the $\Delta^{95}$Mo offset between the NC and CC reservoirs (Fig. 2), and so they do not argue against the presence of an NC-CC dichotomy. Rather, they point towards small secondary Mo isotope variability within the CC reservoir. It is unlikely that this variability is related to processes on the CC parent bodies (Sanders and Scott, 2021), because the large-scale melting and metal segregation on iron meteorites parent bodies will have homogenized any Mo isotope heterogeneity that may have existed within the parent body. Instead, the $\Delta^{95}$Mo variability is more easily understood as reflecting variable abundances of isotopically distinct nebular components within individual parent bodies. This interpretation is developed in more detail in the following section.

### 4.2 Origin of *r*-process variability among CC meteorites

Some recent models have explained the isotopic variations among carbonaceous chondrites by variable mixing of isotopically distinct chondritic components, namely, CI chondrite-like matrix, refractory inclusions (CAIs/AOAs), and chondrules (Alexander, 2019a; Hellmann et al., 2020, 2023). This interpretation is partly based on the observation that the isotopic compositions of carbonaceous chondrites define linear trends when plotted against the matrix mass fractions inferred for each chondrite, indicating that these isotopic variations can be accounted for by binary mixing. For example, Fig. 7 shows that the $\Delta^{17}$O variations among carbonaceous chondrites correlate with the fraction of matrix. The metal-rich chondrites (CR, CH, CB) plot off this correlation, probably because they contain an isotopically distinct chondrule component (Van Kooten et al., 2020; Marrocchi et al., 2022; Hellmann et al., 2023). Unfortunately, available $\Delta^{95}$Mo data for carbonaceous chondrites are too imprecise to assess whether a similar correlation exists for Mo. However, we find a good correlation between $\Delta^{95}$Mo and $\Delta^{17}$O for several CC meteorites, including some achondrites and iron meteorites for which more precise Mo isotope data are available (Fig. 7 and Table S6). This correlation suggests that the $\Delta^{95}$Mo variations among the CC meteorites, like the variations in $\Delta^{17}$O, are also caused by variable abundances of isotopically distinct chondrite components. To assess this idea more quantitatively, we calculated mixing lines between CI-like matrix and chondrules or CAIs (Fig. 8 and Table S7). For the chondrule component, we take the average composition of pooled chondrule separates from Allende (Budde et al., 2016a). For CAIs we used two different CAI populations with distinct Mo isotopic compositions (Brennecka et al., 2020). Specifically, Brennecka et al. (2020) found that group II CAIs have lower Mo concentrations and $\Delta^{95}$Mo values than CAIs with an unfractionated REE pattern. Our calculations show that mixing between CI-like matrix and a chondrule component can account for the correlated Mo and O isotopic variations observed among CC meteorites (Fig. 8). Addition of group II CAIs to a CI-like matrix results in a mixing line with a slope that is too shallow, while addition of unfractionated CAIs, owing to their high Mo/O ratio, results in a steeper mixing line. Overall, these calculations show that addition of ~5 wt% of unfractionated CAIs is sufficient to account for the observed $\Delta^{95}$Mo variability (Fig. 8), consistent with the ~4 vol% of CAIs present in carbonaceous chondrites (Rubin, 2018). As such, it is conceivable that while the O isotope variations reflect mixing of chondrules and matrix, the Mo isotope variations reflect mixing between CAIs and matrix. This situation would be similar to the correlated $^{50}$Ti-$^{54}$Cr isotope variations among carbonaceous chondrites, which probably reflect coupled abundance variations of CAI/chondrules with CI-like matrix (Hellmann et al., 2023). Determining whether CAIs or chondrules are the main carrier of $\Delta^{95}$Mo variations in carbonaceous chondrites will require better knowledge of the Mo isotope composition of the chondrule component. For instance, the average chondrule composition plotted in Fig. 8 is for pooled chondrule separates of the CV chondrite Allende, and it is well possible that these samples already contain some CAI material incorporated into the chondrules during their formation. In this case, CAIs would be the main carrier responsible for the $\Delta^{95}$Mo variations. Either way, our results show that mixing among CI-like matrix, a chondrule component, and CAIs can readily account for the correlated Mo-O isotope variations observed among CC meteorites, including not only chondrites but also achondrites and iron meteorites. Of note, most of the ungrouped CC irons have relatively low $\Delta^{95}$Mo values of ~18 (Fig. 6 and Table 3), suggesting that they are relatively matrix-rich (i.e., poor in chondrules and refractory inclusions), unlike most of the major CC iron groups. Based on the modeled highly siderophile elements (HSE) content of their parental melts, it has been argued that most CC iron meteorite parent bodies accreted between 9 and 26 wt% (IVB) of CAIs (Zhang et al., 2022a, 2024), much higher than the <5 wt% variation inferred above. This mismatch cannot reflect the preferential incorporation of a larger fraction of group II CAIs with low Mo contents, because these CAIs are also not enriched in the HSEs (Mason and Taylor, 1982). Thus, either the iron meteorite parent bodies accreted the HSEs from a yet unsampled component, which is rich in refractory siderophile elements but depleted in Mo, or the HSE contents of these bodies are lower than estimated. For instance, the inferred HSE contents of the iron meteorite's parental melts are strongly controlled by the estimated S content of the melt, which varies between different studies. In fact, the higher S contents for CC iron meteorite cores inferred in some studies (Hilton et al., 2022; Chiappe et al., 2023b) lead to significantly lower HSE abundances, alleviating the need to invoke the accretion of large fractions of CAIs.

In summary, the observation that CC-type chondrites, achondrites, and iron meteorites cover the same range of correlated Mo-O isotope variations suggests strongly that their parent bodies accreted from the same nebular components. This in turn implies that the component mixtures found in carbonaceous chondrites, which formed 2–4 Ma after solar system formation, were already present in the disk at the time when the CC iron meteorite parent bodies accreted less than 1 Ma after CAI formation (Kruijer et al., 2017; Spitzer et al., 2021). Although the exact formation location of CC bodies remains unknown, it is commonly assumed to be beyond the water snow line (Wood, 2005; Warren, 2011; Alexander, 2019a; Lichtenberg et al., 2021; Spitzer et al., 2021; Grewal et al., 2024). Astrophysical models predict the rapid outward transport of refractory materials after their formation close to the Sun due to turbulent diffusion and outward motion of the gas (Yang and Ciesla, 2012; Desch et al., 2018). As such, refractory inclusions were likely present in the outer disk very early on. Similarly, CI-like dust has been proposed to be ubiquitous throughout the outer disk (Van Kooten et al., 2021; Yap and Tissot, 2023; Spitzer et al., 2024). The presence of a chondrule component among the precursor materials of the CC irons may appear more surprising because the majority of the chondrules themselves are thought to have formed more than ~2 Ma after CAI formation (Villeneuve et al., 2009; Hutcheon et al., 2009; Fukuda et al., 2022; Piralla et al., 2023). However, relict chondrules have been identified in some primitive achondrites, suggesting that an earlier generation of chondrules may have existed in the disk (Tomkins et al., 2020; Ma et al., 2022). Moreover, it is important to recognize that the nucleosynthetic isotope anomalies cannot distinguish between chondrules themselves and their precursors. Thus, the early presence of a chondrule component in the accretion region of CC iron meteorite parent bodies does not imply that chondrules themselves were present.

### 4.3 Hf-W model ages and the lack of CC-type non-magmatic iron meteorites

The Hf-W model ages provide insight into whether the dominant heat source for metal melting on the parent bodies was internal or external. Heating by radioactive decay of short-lived $^{26}Al$ led to melting and differentiation of planetesimals formed within the first ~1.5 Ma of the solar system (Hevey and Sanders, 2006). Thermal modeling shows that bodies accreted within this timeframe will undergo core formation within the first ~3 Ma (e.g., Kruijer et al., 2014), consistent with the Hf-W model ages of magmatic iron meteorite groups from both the NC and CC reservoirs (Kruijer et al., 2014b, 2017; Tornabene et al., 2020; Spitzer et al., 2021; Tornabene et al., 2023; Chiappe et al., 2023b). By contrast, younger Hf-W model ages are indicative of an external heat source, which most likely is impact heating. Such younger model ages are found for the non-magmatic irons, such as groups IAB and IIE (Worsham et al., 2017; Hunt et al., 2018; Kruijer and Kleine, 2019; Hilton and Walker, 2020).

Comparison of the Hf-W model ages of the iron meteorite groups to those of the ungrouped irons of this study (Fig. 4) shows the same general systematics, but also reveals some notable differences. First, while the vast majority of the CC irons display Hf-W model ages of ~3.3 Ma after CAI formation, some of the ungrouped irons have older ages of ~1.5 Ma after CAIs (Fig. 4). These older ages overlap with model ages of several of the NC magmatic irons, including some ungrouped NC irons. Thus, unlike the magmatic groups, the ungrouped irons show that core formation ages overlap for the NC and the CC irons. Second, in contrast to the relatively uniform Hf-W model ages of CC irons, the ungrouped NC iron meteorites exhibit a larger range of ages (Fig. 4). Five samples have model ages within ~2 Ma after CAI formation, while four ungrouped NC irons (Cambria, Butler, Guin, and NWA 859) display younger ages of up ~7 Ma after CAIs (Fig. 4). This distribution of ages is similar to that seen for the NC iron meteorite groups, where the magmatic groups have Hf-W model ages of <3 Ma (Kruijer et al., 2017; Spitzer et al., 2021; Tornabene et al., 2023; Chiappe et al., 2023b), while the non-magmatic IAB and IIE irons have younger ages of around ~7 Ma and later (Fig. 4) (Worsham et al., 2017; Kruijer and Kleine, 2019; Hilton and Walker, 2020). As noted above, these younger model ages are indicative of impact heating and melting, consistent with the idea that the non-magmatic irons formed by impact-related processes (Wasson and Wang, 1986; Wasson and Kallemeyn, 2002). A corollary of this interpretation is that the ungrouped NC irons having younger Hf-W model ages are also non-magmatic in origin. According to Wasson (2013), an Ir/Au ratio far from chondritic (Ir/Au <1 and >9) and Au contents ≥0.3 µg/g are indicative of a magmatic origin of an iron meteorite. However, Rubin and Zhang (2023) showed that this criterion may not apply in cases where impact melting of the iron itself occurred. Of the four potentially non-magmatic ungrouped NC irons of this study, only Guin's Ir/Au ratio of 4.3 is within the expected range of values for non-magmatic irons, while Cambria, Butler, and NWA 859 all have Ir/Au ratios just below 1 (0.17–0.63; Table 1). Guin also is petrographically similar to the non-magmatic IIE irons and contains silicate inclusions, another distinctive characteristic of non-magmatic irons (Rubin et al., 1986), while no silicate inclusions have yet been identified for the other three samples. However, Butler and NWA 859 are chemically related but do not show magmatic trends and are thus likely of impact origin (Wasson, 2011). Nevertheless, while there is no supporting chemical or petrographic evidence for a non-magmatic origin of Cambria, their young Hf-W model ages indicate a late metal-silicate fractionation event (or events), inconsistent with a purely magmatic iron-like origin for these samples. We therefore interpret Guin, Cambria, Butler, and NWA 859 as being non-magmatic in origin.

A key observation from the global set of Hf-W model ages for iron meteorites is that younger ages, indicative of a non-magmatic origin, appear to be exclusive to the NC reservoir (Fig. 4). Given the large number of both NC and CC parent bodies for which Hf-W model ages are available, it seems unlikely that this simply reflects a sampling bias (i.e., we have not yet sampled non-magmatic CC irons). In total we have Hf-W model ages for 24 CC iron parent bodies (5 groups, the South Bryon Trio, and 18 individual ungrouped irons), of which none seem to be non-magmatic in origin. For the NC irons, we have Hf-W model ages for 16 parent bodies (7 groups, of which the IABs derive from at least two distinct parent bodies, 7 individual ungrouped irons plus paired Butler and NWA 859), of which 5 are non-magmatic in origin. These numbers may slightly change as some of the ungrouped irons may be paired, but this is unlikely to change the overall picture, namely that there are no non-magmatic CC irons, while about 30% of the NC iron parent bodies have a non-magmatic origin. Moreover, with the sole excpetion of Nedagolla, there is no evidence for the incorporation of CC material into the non-magmatic NC irons, indicating that the collisions involved in their formation occurred predominantly among NC bodies.

It is important to recognize that the Hf-W model ages for non-magmatic irons do not date an event per se, but merely provide a minimum age for the last metal-silicate equilibration event (Worsham et al., 2017; Kruijer and Kleine, 2019). As such, the Hf-W model ages indicate that impacts onto the non-magmatic iron parent bodies occurred later than ~7 Ma after CAI formation, i.e., after gas removal from the solar nebula at ~4–5 Ma (Fu et al., 2014). NC and CC meteorites are thought to have formed inside and outside of Jupiter's orbit, respectively, and were mixed only later during the growth and gas-driven migration of the giant planets, which scattered CC bodies into the inner disk and implanted them into the asteroid belt (Raymond and Izidoro, 2017). Consequently, by 4–5 Ma after CAI formation, NC and CC bodies were both present in the inner disk. As such, the exclusive occurrence of non-magmatic NC irons with evidence for impact heating at >7 Ma cannot be linked to distinct formation locations in the disk, where the intensity of impact heating may have been different. Instead, since by this time NC and CC bodies both were present in the inner disk, this difference more likely reflects different material properties of the NC and CC parent bodies.

Carbonaceous chondrites typically have porosities of ~20 % or higher, while ordinary chondrites have porosities of just under 10 % (Consolmagno et al., 2008). Moreover, the bulk porosities of some C-type asteroids, the likely parent bodies of the carbonaceous chondrites, are even higher (Britt et al., 2002). Importantly, impact heating is strongly reduced during low-velocity collisions of porous planetesimals (De Niem et al., 2018), and so the absence of non-magmatic CC irons likely reflects the more porous and fragile nature of carbonaceous chondrite-like parent bodies compared to their NC counterparts. In other words, unlike for NC bodies, collisions involving porous CC bodies did not result in significant production of impact-generated melt. A corollary of this observation is that the non-magmatic NC irons formed by collisional heating of previously unmelted (i.e., chondritic) NC planetesimals, because otherwise the distinct material properties of chondritic CC and NC bodies could not be cause of the exclusive occurrence of NC non-magmatic irons.

### 4.4 Formation conditions and chemical diversity among NC and CC iron meteorite parent bodies

The association of the ungrouped irons to either the NC or CC reservoir provides genetic information for a large number of chemically diverse parent bodies, and so allows assessing as to whether the formation conditions and chemical characteristics of NC and CC bodies were systematically different. Such differences might be expected, as the distinct genetics of these bodies imply that they formed in different regions of the disk. Importantly, including not only iron meteorite groups but also ungrouped samples in such a comparison more than doubles the number of parent bodies for which genetic and chemical information is available.

*4.4.1 Conditions of core formation*

Based on available data for the major groups of magmatic irons, it was suggested that NC and CC irons are characterized by systematically different chemical compositions, such as their median Ni or refractory siderophile element (RSE) content (Rubin, 2018), which collectively may be indicative of a higher oxidation state of CC parent bodies (Rubin, 2018; Tornabene et al., 2020; Spitzer et al., 2021; Hilton et al., 2022). For the magmatic iron meteorite groups, Grewal et al. (2024) have used the chemical composition derived for each core to deduce the conditions of core formation and in particular the oxygen fugacities prevailing during metal-silicate separation. These authors found that for all irons, including the NC and CC groups, the oxygen fugacity during core formation was between -1 to -3 relative to the iron-wüstite (ΔIW) buffer. To assess whether the larger number of parent bodies sampled by the ungrouped irons show similar systematics, we followed the approach of Grewal et al. (2024), but with some important modifications (Table S8; see Appendix for details). Unlike in the previous study, we did not rely on HSE concentrations for estimating core mass fractions, but used Ni and Co instead, which also nearly quantitatively partition into the core. This approach has two advantages. First, estimates for the parent melt concentrations for the HSEs are strongly dependent on the inferred S content of the bulk core, which varies among different studies and thus leads to different estimated core mass fractions (Hilton et al., 2022; Zhang et al., 2022a). For instance, the different estimates for the S content of the IID core of 0.5 (Zhang et al., 2022a) and 10 wt.% (Hilton et al., 2022) would lead to core mass fractions of 7 and 15 wt.%, respectively. Second, the strongly fractionated HSE pattern estimated for the IVB core has been interpreted to reflect accretion of an ultrarefractory component strongly enriched in HSEs (Campbell and Humayun, 2005), in which case assuming a chondritic bulk HSE compositon of the IVB parent body prior to core formation is inappropriate and leads to core size estimates that are too small. As a result, whereas Grewal et al. (2024) based on HSE systematics derived a core mass fraction for the IVB irons of only 4%, our approach utilizing

Ni and Co yields a higher value of 8%. While this difference may appear small, it significantly changes the inferred oxidation state, from ΔIW = -2.0 for the 4% core mass fraction to ΔIW = -1.3 for an 8% core mass fraction. Importantly, this latter value is more consistent with the ΔIW values inferred for other CC cores (see below).

Estimating the oxidation states for the ungrouped irons is more difficult, as they represent individual samples and so the bulk compositions of their parent cores cannot be reconstructed by fractional crystallization modeling of a co-genetic suite of samples. It thus is necessary to make additional assumptions, as follows. The S content of the parental melt was inferred using the correlation between modeled S content and measured Ga/Ni and Ge/Ni ratios observed for the magmatic iron groups (Fig. S1). For Ni, we use the measured Ni content of the ungrouped irons to estimate their parental melt's Ni abundance following the same approach as outlined for the major iron meteorite groups (see supplementary text A4). This simplification is reasonable because Ni is not strongly fractionated during core crystallization (Scott and Wasson, 1975), and because its partitioning behavior ($D_{Ni}$) is relatively constant across a wide range of S contents (Chabot et al., 2017). For example, a S content of 1 wt% (e.g. IVB irons) results in a $D_{Ni}$ value of 0.88, whereas a S content of 18 wt% (e.g. IC irons) in a $D_{Ni}$ value of 1.12. As such, the inferred Ni contents of the parental melts are within 0–2 wt% of the least evolved member, and so even if some of the ungrouped irons do not reflect early crystallized samples, the bulk Ni contents of their parental cores can nevertheless be estimated this way. We then used the Ni contents of the ungrouped irons to estimate the Co contents and core mass fractions based on the correlations observed among the magmatic iron meteorite groups (Fig. S1). Lastly, for the P content, we use the average of the magmatic NC and CC iron meteorite groups, respectively, because the inferred oxidation state is only marginally influenced by the adopted P content of the parental melts, but mainly a function of their Ni and S contents. With these assumptions, mantle FeO concentrations for the parent bodies of the ungrouped irons can be calculated analogously to the iron meteorite groups.

Our results show that the oxygen fugacities inferred for the parent bodies of the ungrouped irons range from ΔIW values of approximately -1 to -2.5, similar to the range observed among the major iron groups (Fig. 9). The CC irons display a distinct peak at ΔIW = -1.4 (including the revised value for the IVB irons), whereas the distribution of the NC irons is broader. This is at least partly due to the ungrouped NC irons, for some of which we, owing to their higher Ni contents (>9 wt%), infer somewhat more oxidizing conditions than for the NC iron groups. Nevertheless, the ungrouped irons appear to be characterized by similar oxidation states as their counterparts from the iron groups. Importantly, although the ΔIW values of the NC and CC irons overlap, their mean values are distinct. A two-tailed unequal variances t-test reveals that the difference between the means of the magmatic NC and CC irons is statistically significant at the 95% confidence limit (p-value of 0.0145, Fig. 9), indicating that CC irons formed at somewhat more oxidizing conditions than the NC irons.

*4.4.2 Volatile element depletion*

As is evident from variations in the $(Ge/Ni)_{CI}$ and $(Ga/Ni)_{CI}$ ratios, iron meteorites are strongly and variably depleted in moderately volatile elements such as Ge and Ga. These depletions are used for classification of the iron meteorites into distinct chemical groups, but also extend to the ungrouped irons, which show a similar range of depletions (Fig. 10). Based on these depletions it has been suggested that some strongly depleted ungrouped irons may be related to the IVA or IVB irons (Buchwald, 1975; Massart et al., 1981). However, the Mo isotopic data of this study show that in most cases this cannot be the case, because these ungrouped irons have isotopic compositions distinct from the IVA and IVB irons.

A key observation from the comparison of either NC or CC genetics to the degree of moderately volatile element depletion is that there is a clear tendency for more volatile-depleted ungrouped irons to be of CC origin. This observation may seem surprising, because the CC reservoir is in general expected to be characterized by stronger volatile enrichments than the NC reservoir. However, the volatile element depletions observed among the iron meteorites are orders of magnitude larger than those seen among either NC or CC chondrites, and so these depletions are unlikely to be inherited from a parent body's precursor material. Instead, they more likely reflect secondary losses during differentiation and impact disruption of the iron meteorite parent bodies. For instance, based on metallographic cooling rates (Yang et al., 2007) and Pd-Ag isotope systematics (Horan et al., 2012; Matthes et al., 2018), it has been proposed that the volatile-poor nature of the IVA parent body reflects catastrophic impact disruption that lead to near complete stripping of the silicate mantle, exposing the still molten core and facilitating efficient degassing. However, if degassing after collisional mantle stripping is the dominant process causing moderately volatile element depletion among iron meteorites, there is no obvious reason why this depletion would be much stronger for CC compared to NC irons. This is because it should no longer matter for a fully differentiated object if it is of NC or CC heritage. Based on the C and S contents inferred for parental cores of magmatic iron meteorite groups, Hirschmann et al. (2021) argued that the volatile element depletion did not occur by degassing of exposed molten cores, but happened earlier, during surface degassing from silicate melts. However, it is unclear why this process would have been more effective on CC than NC parent bodies.

# 5 Conclusions

This study comprises the largest data set of Mo and W isotopic data for ungrouped iron meteorites and provides genetic and chronological information for 22 previously uninvestigated parent bodies, which can be compared to existing data for 35 NC and CC meteorite groups. The Mo isotopic signatures of the ungrouped irons cover the same range of anomalies as the iron meteorite groups and demonstrate that the ungrouped irons can be assigned to either an NC or a CC heritage. Likewise, Hf-W model ages are also similar for ungrouped irons and the iron meteorite groups, collectively indicating that core formation in iron meteorite parent bodies occurred within ~3 Ma after CAI formation. Together, these observations confirm that the isotopic dichotomy between NC and CC meteorites is a fundamental feature of early-formed planetary objects and was established within ~1 Ma of solar system history.

The new data for ungrouped NC irons all plot on the previously defined NC-line in the $\varepsilon^{94}$Mo–$\varepsilon^{95}$Mo diagram and demonstrate that all 41 NC meteorite parent bodies for which Mo isotope data are available plot on a single, precisely defined NC-line. By contrast, several ungrouped CC irons scatter slightly below the CC-line, indicative of small variations in the distribution of *r*-process Mo among the CC meteorites. Correlation of these *r*-process Mo isotope variations with O isotope anomalies suggests that these variations reflect variable mixtures of CI chondrite-like matrix, chondrule precursors, and refractory inclusions. A corollary of this observation is that early-formed CC bodies (<1 Ma after CAIs) formed from the same principal materials as the later-formed CC chondrites (2–4 Ma after CAIs), which in turn implies that these materials have been stored in the outer disk for several million years.

The new data for the ungrouped irons allow a more general comparison between NC and CC irons. Here, three observations stand out. First, non-magmatic iron meteorites appear to be exclusive to the NC reservoir, most likely because the more fragile and porous nature of CC parent bodies disfavored the formation of large impact melt pools. Second, although overall similar, the conditions of core formation were slightly more oxidizing in CC than in NC iron parent bodies. Finally, objects with strong depletions in moderately volatile elements are more abundant among CC than NC bodies, for reasons that remain unclear at present.

## Data Availability

Data are available through Mendeley Data at https://data.mendeley.com/datasets/dgsv89g9n4/2.

## CRediT author statement

Fridolin Spitzer: Investigation, Formal analysis, Visualization, Data Curation, Writing – Original draft preparation, Reviewing and Editing. Christoph Burkhardt: Conceptualization, Funding acquisition, Supervision, Writing – Reviewing and Editing. Thomas S. Kruijer: Writing – Reviewing and Editing. Thorsten Kleine: Conceptualization, Funding acquisition, Supervision, Writing – Reviewing and Editing.

## Acknowledgments

We thank the Field Museum Chicago, the Natural History Museum London, the Smithsonian Institution, the Max Planck Institute for Nuclear Physics in Heidelberg, and the Institut für Planetologie in Münster for generously providing meteorite samples for this study. We are thankful for constructive reviews by A. Rubin and an anonymous reviewer. This work was funded by the Deutsche Forschungsgemeinschaft (DFG, German Research Foundation) (Project- ID 263649064-TRR170). This is TRR publication #223.

## Appendix A. Supplementary Material

Supplementary material related to this article includes a detailed description of the chemical separation and measurement protocols for W, Mo, and Pt isotope analyses. It also provides more background on calculating the oxygen fugacities during core formation in the iron meteorite parent bodies and presents an update on the correction of cosmic-ray exposure effects on Mo isotopes. Furthermore, it contains several tables with compiled chemical and isotopic data relevant to the discussion of this study.

# Tables

**Table 1.** Sample characteristics of the ungrouped iron meteorites.

| Sample | Source | Item No | Weight (g) | Ir/Ni ($x10^4$) | Ir/Au | $\Delta^{17}O$ |
|---|---|---|---|---|---|---|
| Babb's Mill (Troost's Iron) | NHM London | BM.18490 | 0.57776 | 1.70 | 33 | -3.6 |
| Butler | Field Museum Chicago | ME 96 #4 | 0.51188 | 0.0623 | 0.17 | |
| Cambria | NHM London | BM.19005 | 0.48256 | 0.0839 | 0.39 | |
| Grand Rapids | NHM London | BM.68724 | 0.53816 | 1.83 | 23 | |
| Nedagolla | NHM London | BM.1985,M268 | 0.56057 | 0.867 | 24 | |
| Piñon | NHM London | BM.1959,911 | 0.43148 | 0.965 | | |
| Reed City | University of Münster | | 0.42581 | 5.79 | 35 | |
| Tucson | Field Museum Chicago | ME 858 #8 | 0.53882 | 0.222 | 2.1 | -2.51 |
| Zacatecas (1792) | Field Museum Chicago | ME 28 #1 | 0.53564 | 0.357 | 2.9 | |
| Guin | University of Münster | | 0.64044 | 0.719 | 4.3 | 1.13 |
| Mbosi | Field Museum Chicago | ME 2606 #3 | 0.63948 | 0.751 | 17 | -2.84 |
| New Baltimore | Smithsonian Institution | USNM 710 | 0.55715 | 1.85 | 34 | |
| Nordheim | Smithsonian Institution | USNM 3190 | 0.48178 | 0.945 | 29 | |
| NWA 6932 | Smithsonian Institution | USNM 7612 | 0.51297 | 0.641 | 10 | |
| Santiago Papasquiero | Smithsonian Institution | USNM 2603 | 0.57925 | 0.535 | 9.1 | |
| Tishomingo | Smithsonian Institution | USNM 5862 | 0.52945 | 0.542 | 121 | -0.15 |
| Washington County | MPI for Nuclear Physics Heidelberg | MPK 3078A | 0.57549 | 0.0067 | 0.052 | |
| ALHA 77255 | University of Münster | | 0.63645 | 0.805 | | -0.48 |
| EET 83230 | University of Münster | | 0.51468 | 0.0052 | 0.023 | 1.13 |
| Guffey | Smithsonian Institution | USNM 4832 | 0.62807 | 0.503 | 41 | |
| Hammond | Smithsonian Institution | USNM 471 | 0.63224 | 0.0122 | 0.07 | |
| ILD 83500 | University of Münster | | 0.55395 | 0.409 | 4.4 | -3.7 |
| Illinois Gulch | NHM London | BM.84785 | 0.50438 | 0.454 | | |
| La Caille | Field Museum Chicago | ME 902 #2 | 0.56304 | 1.00 | 13 | |
| N'Goureyma | Field Museum Chicago | ME 978 #7 | 0.55444 | 0.0626 | 2.4 | |
| NWA 859 | University of Münster | | 0.48863 | 0.243 | 0.63 | |

Ir/Ni ratio as a proxy for the refractory siderophile element (RSE) content of a sample after Rubin (2018). Ir/Au ratio as a proxy for a sample's magmatic (Ir/Au <1 and >9) or non-magmatic origin (1≤ Ir/Au ≤9) after Wasson (2013). The $\Delta^{17}O$ values are from Clayton and Mayeda (1996), McCoy et al. (2011, 2019), and Corrigan et al. (2022).

**Table 2.** Pt isotope data for the ungrouped iron meteorites.

| Sample | N (Pt-IC) | $\varepsilon^{192}$Pt (± 2σ) | $\varepsilon^{194}$Pt (± 2σ) | $\varepsilon^{198}$Pt (± 2σ) | $\varepsilon^{192}$Pt (± 2σ) | $\varepsilon^{194}$Pt (± 2σ) | $\varepsilon^{196}$Pt (± 2σ) | $\varepsilon^{196}Pt_{nuc.-corr.}$ (± 2σ) |
|---|---|---|---|---|---|---|---|---|
| | | normalized to $^{196}$Pt/$^{195}$Pt = 0.7464 ('6/5') | | | normalized to $^{198}$Pt/$^{195}$Pt = 0.2145 ('8/5') | | | |
| **NC meteorites** | | | | | | | | |
| Butler | 10 | 2.29 ± 0.41 | 0.60 ± 0.04 | -1.02 ± 0.06 | 1.29 ± 0.34 | 0.26 ± 0.03 | 0.34 ± 0.02 | 0.40 ± 0.02 |
| Cambria | 1 | 11.28 ± 1.21 | 1.05 ± 0.17 | -1.48 ± 0.21 | 9.80 ± 1.18 | 0.55 ± 0.12 | 0.49 ± 0.07 | 0.55 ± 0.07 |
| EET 83230 | 3 | -0.48 ± 1.21 | -0.20 ± 0.17 | 0.13 ± 0.21 | -0.38 ± 1.18 | -0.15 ± 0.12 | -0.04 ± 0.07 | 0.01 ± 0.07 |
| Guin | 7 | 2.86 ± 0.70 | 0.13 ± 0.04 | -0.14 ± 0.06 | 2.75 ± 0.73 | 0.08 ± 0.05 | 0.05 ± 0.02 | 0.11 ± 0.02 |
| NWA 859 | 15 | 1.72 ± 0.28 | 0.47 ± 0.02 | -0.80 ± 0.06 | 0.87 ± 0.26 | 0.20 ± 0.01 | 0.27 ± 0.02 | 0.33 ± 0.02 |
| Reed City | 11 | 6.00 ± 0.34 | 0.19 ± 0.03 | -0.13 ± 0.06 | 5.91 ± 0.35 | 0.15 ± 0.03 | 0.04 ± 0.02 | 0.10 ± 0.02 |
| Santiago Papasquiero | 9 | -0.28 ± 0.61 | 0.11 ± 0.16 | -0.15 ± 0.20 | -0.42 ± 0.52 | 0.07 ± 0.12 | 0.05 ± 0.07 | 0.11 ± 0.07 |
| Washington County | 6 | -0.09 ± 0.57 | 0.06 ± 0.13 | -0.18 ± 0.16 | -0.21 ± 0.58 | 0.00 ± 0.08 | 0.06 ± 0.05 | 0.12 ± 0.05 |
| Zacatecas (1792) | 7 | 0.43 ± 0.36 | 0.11 ± 0.07 | -0.11 ± 0.15 | 0.33 ± 0.33 | 0.07 ± 0.05 | 0.03 ± 0.05 | 0.09 ± 0.05 |
| **CC meteorites** | | | | | | | | |
| ALHA 77255 | 10 | 15.08 ± 0.42 | 0.63 ± 0.06 | -0.67 ± 0.07 | 14.38 ± 0.43 | 0.41 ± 0.04 | 0.22 ± 0.02 | 0.28 ± 0.03 |
| Babb's Mill (Troost's Iron)[a] | 31 | -0.79 ± 0.16 | -0.11 ± 0.02 | 0.17 ± 0.04 | -0.65 ± 0.17 | -0.06 ± 0.02 | -0.06 ± 0.01 | 0.00 ± 0.02 |
| Grand Rapids | 9 | -0.06 ± 0.46 | 0.02 ± 0.04 | 0.03 ± 0.08 | 0.01 ± 0.52 | 0.04 ± 0.03 | -0.01 ± 0.03 | 0.05 ± 0.03 |
| Guffey[a] | 16 | -0.52 ± 0.32 | -0.13 ± 0.05 | 0.20 ± 0.07 | -0.33 ± 0.31 | -0.07 ± 0.03 | -0.07 ± 0.02 | -0.01 ± 0.03 |
| Hammond[a] | 5 | -0.95 ± 0.58 | -0.10 ± 0.09 | 0.20 ± 0.20 | -0.79 ± 0.62 | -0.03 ± 0.05 | -0.07 ± 0.07 | -0.01 ± 0.07 |
| ILD 83500[a] | 11 | -0.70 ± 0.33 | -0.04 ± 0.07 | 0.09 ± 0.09 | -0.64 ± 0.35 | -0.01 ± 0.04 | -0.03 ± 0.03 | 0.03 ± 0.03 |
| Illinois Gulch | 5 | 2.99 ± 0.79 | 0.08 ± 0.06 | -0.07 ± 0.10 | 2.99 ± 0.63 | 0.06 ± 0.05 | 0.02 ± 0.03 | 0.08 ± 0.04 |
| La Caille | 5 | 13.61 ± 0.72 | 0.63 ± 0.09 | -0.71 ± 0.17 | 12.87 ± 0.70 | 0.39 ± 0.04 | 0.24 ± 0.06 | 0.30 ± 0.06 |
| Mbosi | 4 | -0.19 ± 0.59 | 0.06 ± 0.15 | -0.07 ± 0.12 | -0.21 ± 0.50 | 0.06 ± 0.15 | 0.02 ± 0.04 | 0.08 ± 0.04 |
| New Baltimore | 5 | 4.66 ± 0.99 | 0.12 ± 0.07 | -0.09 ± 0.12 | 4.51 ± 0.92 | 0.13 ± 0.07 | 0.03 ± 0.04 | 0.09 ± 0.04 |
| N'Goureyma | 1 | 22.97 ± 1.21 | 1.00 ± 0.17 | -1.14 ± 0.21 | 21.84 ± 1.18 | 0.62 ± 0.12 | 0.38 ± 0.07 | 0.44 ± 0.07 |
| Nordheim | 8 | 2.73 ± 0.38 | 0.11 ± 0.04 | -0.09 ± 0.12 | 2.69 ± 0.34 | 0.08 ± 0.02 | 0.03 ± 0.04 | 0.09 ± 0.04 |
| NWA 6932 | 9 | -0.20 ± 0.47 | 0.07 ± 0.05 | -0.05 ± 0.10 | -0.25 ± 0.36 | 0.06 ± 0.04 | 0.02 ± 0.03 | 0.07 ± 0.04 |
| Piñon | 12 | 12.19 ± 0.25 | 0.70 ± 0.03 | -0.88 ± 0.05 | 11.30 ± 0.24 | 0.40 ± 0.03 | 0.29 ± 0.02 | 0.35 ± 0.02 |
| Tishomingo | 10 | 5.35 ± 0.43 | 0.20 ± 0.04 | -0.23 ± 0.06 | 5.14 ± 0.43 | 0.13 ± 0.04 | 0.08 ± 0.02 | 0.14 ± 0.02 |
| Tucson | 2 | 7.10 ± 1.21 | 0.55 ± 0.17 | -0.61 ± 0.21 | 6.58 ± 1.18 | 0.37 ± 0.12 | 0.20 ± 0.07 | 0.26 ± 0.07 |
| **Mixed composition** | | | | | | | | |
| Nedagolla[b] | 7 | 1.10 ± 0.80 | 0.45 ± 0.11 | -0.46 ± 0.12 | 0.61 ± 0.85 | 0.32 ± 0.05 | 0.16 ± 0.04 | 0.22 ± 0.04 |

Given uncertainties are based on the external reproducibility (2 s.d.) obtained from repeated analyses of the terrestrial standards. For samples with $N \geq 4$, the uncertainties represent Student-t 95% confidence intervals, *i.e.*, $(t_{0.95,N-1} \times \text{s.d.})/\sqrt{N}$. The samples are arranged in alphabetical order within each reservoir. $N$: number of analyses.

[a] (Spitzer et al., 2021)

[b] (Spitzer et al., 2022)

**Table 3.** Mo isotope data for the ungrouped iron meteorites.

| Sample | N | $\varepsilon^{92}Mo_{meas.}$ | $\varepsilon^{94}Mo_{meas.}$ | $\varepsilon^{95}Mo_{meas.}$ | $\varepsilon^{97}Mo_{meas.}$ | $\varepsilon^{100}Mo_{meas.}$ | $\varepsilon^{92}Mo_{CRE.\ corr.}$ | $\varepsilon^{94}Mo_{CRE.\ corr.}$ | $\varepsilon^{95}Mo_{CRE.\ corr.}$ | $\varepsilon^{97}Mo_{CRE.\ corr.}$ | $\varepsilon^{100}Mo_{CRE.\ corr.}$ | $\Delta^{95}Mo$ |
|---|---|---|---|---|---|---|---|---|---|---|---|---|
| | (Mo-IC) | (± 95% CI) | (± 95% CI) | (± 95% CI) | (± 95% CI) | (± 95% CI) | (± 95% CI) | (± 95% CI) | (± 95% CI) | (± 95% CI) | (± 95% CI) | (± 95% CI) |
| **NC meteorites** | | | | | | | | | | | | |
| Butler | 10 | 0.57 ± 0.11 | 0.62 ± 0.07 | 0.13 ± 0.07 | 0.16 ± 0.04 | 0.07 ± 0.09 | 0.77 ± 0.12 | 0.74 ± 0.07 | 0.26 ± 0.07 | 0.20 ± 0.04 | 0.01 ± 0.09 | -19 ± 8 |
| Cambria | 11 | 0.97 ± 0.08 | 0.88 ± 0.05 | 0.30 ± 0.04 | 0.20 ± 0.03 | 0.35 ± 0.04 | 1.24 ± 0.10 | 1.05 ± 0.06 | 0.48 ± 0.06 | 0.25 ± 0.04 | 0.26 ± 0.05 | -15 ± 7 |
| EET 83230 | 9 | 0.84 ± 0.14 | 0.71 ± 0.09 | 0.36 ± 0.05 | 0.19 ± 0.05 | 0.18 ± 0.06 | 0.85 ± 0.15 | 0.72 ± 0.10 | 0.36 ± 0.05 | 0.19 ± 0.05 | 0.18 ± 0.07 | -7 ± 8 |
| Guin | 6 | 0.57 ± 0.17 | 0.52 ± 0.11 | 0.24 ± 0.07 | 0.16 ± 0.07 | 0.18 ± 0.08 | 0.62 ± 0.17 | 0.56 ± 0.11 | 0.26 ± 0.07 | 0.17 ± 0.07 | 0.17 ± 0.08 | -7 ± 9 |
| NWA 859 | 6 | 0.59 ± 0.13 | 0.58 ± 0.04 | 0.19 ± 0.05 | 0.17 ± 0.05 | 0.21 ± 0.12 | 0.74 ± 0.13 | 0.68 ± 0.05 | 0.29 ± 0.05 | 0.20 ± 0.05 | 0.16 ± 0.12 | -11 ± 6 |
| Reed City | 13 | 1.09 ± 0.09 | 0.96 ± 0.06 | 0.43 ± 0.04 | 0.24 ± 0.02 | 0.20 ± 0.05 | 1.13 ± 0.09 | 0.99 ± 0.06 | 0.44 ± 0.05 | 0.25 ± 0.02 | 0.18 ± 0.05 | -15 ± 6 |
| Santiago Papasquiero | 9 | 0.72 ± 0.12 | 0.69 ± 0.05 | 0.39 ± 0.04 | 0.25 ± 0.05 | 0.15 ± 0.09 | 0.77 ± 0.12 | 0.73 ± 0.06 | 0.40 ± 0.05 | 0.26 ± 0.05 | 0.13 ± 0.10 | -3 ± 6 |
| Washington County | 6 | 1.11 ± 0.09 | 1.03 ± 0.07 | 0.47 ± 0.03 | 0.23 ± 0.06 | 0.30 ± 0.09 | 1.17 ± 0.10 | 1.07 ± 0.07 | 0.49 ± 0.04 | 0.24 ± 0.06 | 0.28 ± 0.09 | -14 ± 5 |
| Zacatecas (1792) | 8 | 1.45 ± 0.13 | 1.22 ± 0.08 | 0.59 ± 0.04 | 0.32 ± 0.04 | 0.37 ± 0.05 | 1.50 ± 0.13 | 1.25 ± 0.09 | 0.60 ± 0.05 | 0.33 ± 0.04 | 0.35 ± 0.05 | -14 ± 7 |
| **CC meteorites** | | | | | | | | | | | | |
| ALHA 77255 | 7 | 1.79 ± 0.14 | 1.35 ± 0.07 | 0.96 ± 0.07 | 0.50 ± 0.04 | 0.59 ± 0.09 | 1.93 ± 0.14 | 1.44 ± 0.07 | 1.04 ± 0.07 | 0.52 ± 0.04 | 0.55 ± 0.09 | 19 ± 8 |
| Babb's Mill (Troost's Iron)[a] | 7 | 1.89 ± 0.14 | 1.31 ± 0.10 | 1.07 ± 0.08 | 0.51 ± 0.07 | 0.55 ± 0.10 | 1.89 ± 0.14 | 1.31 ± 0.10 | 1.07 ± 0.08 | 0.51 ± 0.07 | 0.55 ± 0.10 | 29 ± 10 |
| Grand Rapids | 8 | 1.62 ± 0.16 | 1.19 ± 0.10 | 0.94 ± 0.05 | 0.41 ± 0.03 | 0.52 ± 0.07 | 1.65 ± 0.16 | 1.20 ± 0.10 | 0.94 ± 0.05 | 0.41 ± 0.03 | 0.52 ± 0.07 | 22 ± 8 |
| Guffey[a] | 6 | 1.82 ± 0.17 | 1.42 ± 0.11 | 1.03 ± 0.05 | 0.55 ± 0.05 | 0.60 ± 0.10 | 1.82 ± 0.17 | 1.42 ± 0.11 | 1.03 ± 0.05 | 0.55 ± 0.05 | 0.60 ± 0.10 | 18 ± 8 |
| Hammond[a] | 6 | 2.28 ± 0.16 | 1.76 ± 0.09 | 1.29 ± 0.03 | 0.65 ± 0.07 | 0.66 ± 0.08 | 2.28 ± 0.16 | 1.76 ± 0.10 | 1.29 ± 0.04 | 0.65 ± 0.07 | 0.66 ± 0.08 | 24 ± 7 |
| ILD 83500[a] | 6 | 1.71 ± 0.16 | 1.20 ± 0.14 | 1.01 ± 0.08 | 0.53 ± 0.05 | 0.64 ± 0.10 | 1.72 ± 0.16 | 1.21 ± 0.14 | 1.01 ± 0.08 | 0.53 ± 0.05 | 0.64 ± 0.10 | 29 ± 12 |
| Illinois Gulch | 6 | 1.85 ± 0.15 | 1.43 ± 0.07 | 1.03 ± 0.07 | 0.54 ± 0.07 | 0.68 ± 0.08 | 1.89 ± 0.15 | 1.46 ± 0.07 | 1.04 ± 0.07 | 0.55 ± 0.07 | 0.66 ± 0.08 | 17 ± 8 |
| La Caille | 6 | 1.82 ± 0.10 | 1.42 ± 0.05 | 1.00 ± 0.04 | 0.56 ± 0.05 | 0.74 ± 0.06 | 1.96 ± 0.11 | 1.51 ± 0.06 | 1.09 ± 0.05 | 0.58 ± 0.05 | 0.69 ± 0.07 | 18 ± 6 |
| Mbosi | 7 | 1.70 ± 0.11 | 1.25 ± 0.12 | 1.01 ± 0.05 | 0.51 ± 0.06 | 0.49 ± 0.14 | 1.74 ± 0.11 | 1.28 ± 0.12 | 1.02 ± 0.05 | 0.52 ± 0.06 | 0.48 ± 0.14 | 25 ± 9 |
| New Baltimore | 8 | 2.25 ± 0.16 | 1.73 ± 0.13 | 1.19 ± 0.07 | 0.64 ± 0.03 | 0.72 ± 0.07 | 2.29 ± 0.16 | 1.75 ± 0.13 | 1.20 ± 0.07 | 0.65 ± 0.03 | 0.71 ± 0.07 | 16 ± 10 |
| N'Goureyma | 6 | 1.67 ± 0.23 | 1.35 ± 0.09 | 0.94 ± 0.07 | 0.54 ± 0.07 | 0.63 ± 0.12 | 1.88 ± 0.24 | 1.49 ± 0.10 | 1.08 ± 0.08 | 0.57 ± 0.07 | 0.57 ± 0.13 | 19 ± 10 |
| Nordheim | 6 | 2.08 ± 0.20 | 1.55 ± 0.07 | 1.13 ± 0.05 | 0.56 ± 0.02 | 0.61 ± 0.16 | 2.12 ± 0.21 | 1.58 ± 0.08 | 1.14 ± 0.05 | 0.56 ± 0.02 | 0.60 ± 0.16 | 20 ± 7 |
| NWA 6932 | 6 | 1.65 ± 0.26 | 1.18 ± 0.09 | 0.99 ± 0.09 | 0.49 ± 0.05 | 0.53 ± 0.09 | 1.69 ± 0.26 | 1.20 ± 0.09 | 0.99 ± 0.10 | 0.49 ± 0.05 | 0.52 ± 0.09 | 28 ± 11 |
| Piñon | 7 | 1.52 ± 0.08 | 1.06 ± 0.06 | 0.84 ± 0.06 | 0.44 ± 0.05 | 0.65 ± 0.06 | 1.69 ± 0.09 | 1.17 ± 0.07 | 0.95 ± 0.06 | 0.47 ± 0.05 | 0.59 ± 0.06 | 25 ± 8 |
| Tishomingo | 10 | 1.73 ± 0.17 | 1.35 ± 0.09 | 0.97 ± 0.05 | 0.50 ± 0.04 | 0.60 ± 0.05 | 1.80 ± 0.17 | 1.39 ± 0.09 | 1.00 ± 0.05 | 0.51 ± 0.04 | 0.58 ± 0.05 | 17 ± 7 |
| Tucson | 8 | 2.75 ± 0.15 | 2.13 ± 0.07 | 1.49 ± 0.05 | 0.80 ± 0.06 | 1.00 ± 0.07 | 2.88 ± 0.16 | 2.21 ± 0.08 | 1.57 ± 0.06 | 0.82 ± 0.06 | 0.96 ± 0.08 | 25 ± 8 |
| **Mixed composition** | | | | | | | | | | | | |
| Nedagolla[b] | 8 | 0.66 ± 0.05 | 0.54 ± 0.07 | 0.32 ± 0.06 | 0.15 ± 0.04 | 0.14 ± 0.07 | 0.77 ± 0.06 | 0.60 ± 0.08 | 0.38 ± 0.06 | 0.17 ± 0.04 | 0.10 ± 0.07 | 2 ± 8 |

Mo isotope ratios are normalized to $^{98}Mo/^{96}Mo$ = 1.453173. Uncertainties represent the Student-t 95% confidence intervals (95% CI), *i.e.*, $(t_{0.95,N-1} \times s.d.)/\sqrt{N}$. Note that the final uncertainties include all propagated uncertainties induced by the correction for cosmic ray exposure effects ($\varepsilon^{i}Mo_{CRE\text{-}corr.}$). The samples are arranged in alphabetical order within each reservoir. *N*: number of analyses.
[a] (Spitzer et al., 2021)
[b] (Spitzer et al., 2022)

**Table 4.** W isotope data for the ungrouped iron meteorites.

| Sample | N | $\varepsilon^{182}W_{meas.}$ | $\varepsilon^{183}W_{meas.}$ | $\varepsilon^{182}W_{nuc.\ corr.}$ | $\varepsilon^{182}W_{CRE.\ corr.}$ | $\Delta t_{CAI}$ (Ma) | $\varepsilon^{182}W_{meas.}$ | $\varepsilon^{184}W_{meas.}$ | $\varepsilon^{182}W_{nuc.\ corr.}$ | $\varepsilon^{182}W_{CRE.\ corr.}$ | $\Delta t_{CAI}$ (Ma) |
|---|---|---|---|---|---|---|---|---|---|---|---|
| | (W-IC) | (± 2σ) | (± 2σ) | (± 2σ) | (± 2σ) | (± 2σ) | (± 2σ) | (± 2σ) | (± 2σ) | (± 2σ) | (± 2σ) |
| | | normalized to $^{186}W/^{184}W$ = 0.92767 ('6/4') | | | | | normalized to $^{186}W/^{183}W$ = 1.98590 ('6/3') | | | | |
| **NC meteorites** | | | | | | | | | | | |
| Butler | 7 | -3.47 ± 0.06 | 0.01 ± 0.10 | -3.47 ± 0.06 | -2.95 ± 0.07 | 7.2 ± 1.8 | -3.47 ± 0.11 | -0.01 ± 0.07 | -3.47 ± 0.11 | -2.94 ± 0.12 | 7.3 ± 2.4 |
| Cambria | 4 | -3.84 ± 0.08 | 0.01 ± 0.10 | -3.84 ± 0.08 | -3.11 ± 0.13 | 4.6 ± 2.1 | -3.89 ± 0.13 | -0.01 ± 0.07 | -3.89 ± 0.13 | -3.16 ± 0.17 | 3.9 ± 2.4 |
| EET 83230 | 1 | -3.46 ± 0.08 | 0.04 ± 0.10 | -3.46 ± 0.08 | -3.44 ± 0.13 | 0.4 ± 1.2 | -3.56 ± 0.13 | -0.03 ± 0.07 | -3.56 ± 0.13 | -3.54 ± 0.16 | -0.4 ± 1.4 |
| Guin | 7 | -3.12 ± 0.06 | 0.00 ± 0.07 | -3.12 ± 0.06 | -2.98 ± 0.07 | 6.7 ± 1.7 | -3.15 ± 0.10 | 0.00 ± 0.05 | -3.15 ± 0.10 | -3.01 ± 0.10 | 6.1 ± 2.0 |
| NWA 859 | 7 | -3.39 ± 0.04 | 0.03 ± 0.09 | -3.39 ± 0.04 | -2.96 ± 0.05 | 7.1 ± 1.5 | -3.43 ± 0.10 | -0.02 ± 0.06 | -3.43 ± 0.10 | -3.00 ± 0.11 | 6.3 ± 2.0 |
| Reed City | 7 | -3.42 ± 0.04 | 0.04 ± 0.09 | -3.42 ± 0.04 | -3.30 ± 0.05 | 2.1 ± 1.0 | -3.48 ± 0.11 | -0.03 ± 0.06 | -3.48 ± 0.11 | -3.35 ± 0.11 | 1.5 ± 1.5 |
| Santiago Papasquiero | 5 | -3.40 ± 0.09 | -0.04 ± 0.08 | -3.40 ± 0.09 | -3.26 ± 0.13 | 2.6 ± 1.8 | -3.40 ± 0.15 | 0.02 ± 0.06 | -3.40 ± 0.15 | -3.26 ± 0.17 | 2.6 ± 2.3 |
| Washington County | 3 | -3.43 ± 0.08 | 0.06 ± 0.10 | -3.43 ± 0.08 | -3.27 ± 0.11 | 2.4 ± 1.6 | -3.55 ± 0.13 | -0.04 ± 0.07 | -3.55 ± 0.13 | -3.39 ± 0.15 | 1.0 ± 1.8 |
| Zacatecas (1792) | 7 | -3.49 ± 0.04 | 0.00 ± 0.09 | -3.49 ± 0.04 | -3.36 ± 0.08 | 1.3 ± 1.1 | -3.52 ± 0.11 | 0.00 ± 0.06 | -3.52 ± 0.11 | -3.40 ± 0.13 | 1.0 ± 1.6 |
| **CC meteorites** | | | | | | | | | | | |
| ALHA 77255 | 7 | -3.23 ± 0.05 | 0.18 ± 0.08 | -3.48 ± 0.12 | -3.11 ± 0.13 | 3.5 ± 1.5 | -3.51 ± 0.10 | -0.12 ± 0.05 | -3.52 ± 0.10 | -3.15 ± 0.10 | 3.1 ± 1.2 |
| Babb's Mill (Troost's Iron)[a] | 7 | -3.03 ± 0.06 | 0.16 ± 0.08 | -3.26 ± 0.13 | -3.26 ± 0.14 | 2.1 ± 1.4 | -3.28 ± 0.11 | -0.11 ± 0.06 | -3.29 ± 0.11 | -3.29 ± 0.11 | 1.7 ± 1.2 |
| Grand Rapids | 7 | -3.19 ± 0.03 | 0.09 ± 0.07 | -3.32 ± 0.10 | -3.25 ± 0.11 | 2.1 ± 1.2 | -3.35 ± 0.10 | -0.06 ± 0.05 | -3.36 ± 0.10 | -3.29 ± 0.11 | 1.7 ± 1.2 |
| Guffey[a] | 5 | -2.92 ± 0.05 | 0.13 ± 0.08 | -3.10 ± 0.12 | -3.10 ± 0.12 | 3.7 ± 1.4 | -3.13 ± 0.10 | -0.09 ± 0.05 | -3.14 ± 0.10 | -3.14 ± 0.10 | 3.2 ± 1.2 |
| Hammond[a] | 4 | -2.87 ± 0.06 | 0.18 ± 0.11 | -3.13 ± 0.16 | -3.13 ± 0.16 | 3.3 ± 1.8 | -3.15 ± 0.12 | -0.12 ± 0.07 | -3.17 ± 0.12 | -3.17 ± 0.12 | 3.0 ± 1.3 |
| ILD 83500[a] | 5 | -3.09 ± 0.07 | 0.14 ± 0.10 | -3.29 ± 0.15 | -3.29 ± 0.15 | 1.7 ± 1.5 | -3.32 ± 0.11 | -0.09 ± 0.06 | -3.33 ± 0.11 | -3.33 ± 0.11 | 1.4 ± 1.2 |
| Illinois Gulch | 7 | -3.07 ± 0.08 | 0.14 ± 0.10 | -3.26 ± 0.16 | -3.15 ± 0.17 | 3.1 ± 1.9 | -3.29 ± 0.13 | -0.09 ± 0.07 | -3.30 ± 0.13 | -3.19 ± 0.14 | 2.7 ± 1.5 |
| La Caille | 7 | -3.25 ± 0.02 | 0.18 ± 0.07 | -3.50 ± 0.10 | -3.10 ± 0.13 | 3.6 ± 1.5 | -3.52 ± 0.10 | -0.12 ± 0.05 | -3.53 ± 0.10 | -3.14 ± 0.13 | 3.2 ± 1.4 |
| Mbosi | 7 | -3.12 ± 0.03 | 0.12 ± 0.07 | -3.29 ± 0.10 | -3.18 ± 0.11 | 2.8 ± 1.3 | -3.32 ± 0.10 | -0.08 ± 0.05 | -3.32 ± 0.10 | -3.22 ± 0.12 | 2.4 ± 1.3 |
| New Baltimore | 7 | -2.96 ± 0.04 | 0.18 ± 0.08 | -3.22 ± 0.12 | -3.10 ± 0.13 | 3.6 ± 1.5 | -3.24 ± 0.10 | -0.12 ± 0.05 | -3.25 ± 0.10 | -3.14 ± 0.11 | 3.3 ± 1.3 |
| N'Goureyma | 2 | -3.50 ± 0.08 | 0.11 ± 0.10 | -3.65 ± 0.16 | -3.06 ± 0.19 | 4.0 ± 2.2 | -3.67 ± 0.13 | -0.07 ± 0.07 | -3.68 ± 0.13 | -3.10 ± 0.17 | 3.6 ± 1.9 |
| Nordheim | 5 | -2.91 ± 0.02 | 0.21 ± 0.07 | -3.20 ± 0.11 | -3.08 ± 0.12 | 3.8 ± 1.4 | -3.22 ± 0.10 | -0.14 ± 0.05 | -3.23 ± 0.11 | -3.11 ± 0.12 | 3.5 ± 1.4 |
| NWA 6932 | 6 | -3.11 ± 0.06 | 0.09 ± 0.08 | -3.24 ± 0.12 | -3.15 ± 0.13 | 3.2 ± 1.5 | -3.27 ± 0.10 | -0.06 ± 0.05 | -3.28 ± 0.10 | -3.18 ± 0.11 | 2.8 ± 1.3 |

| | | | | | | | | | | |
|---|---|---|---|---|---|---|---|---|---|---|
| Piñon | 7 | -3.50 ± 0.05 | 0.15 ± 0.08 | -3.71 ± 0.12 | -3.25 ± 0.12 | 2.1 ± 1.3 | -3.74 ± 0.11 | -0.10 ± 0.05 | -3.76 ± 0.11 | -3.29 ± 0.11 | 1.7 ± 1.2 |
| Tishomingo | 2 | -2.87 ± 0.08 | 0.26 ± 0.10 | -3.24 ± 0.16 | -3.06 ± 0.17 | 4.0 ± 2.0 | -3.27 ± 0.13 | -0.18 ± 0.07 | -3.28 ± 0.13 | -3.11 ± 0.14 | 3.6 ± 1.6 |
| Tucson | 4 | -3.10 ± 0.08 | 0.18 ± 0.10 | -3.36 ± 0.16 | -3.02 ± 0.19 | 4.6 ± 2.3 | -3.38 ± 0.13 | -0.12 ± 0.07 | -3.40 ± 0.13 | -3.05 ± 0.16 | 4.2 ± 1.9 |
| **Mixed composition** | | | | | | | | | | | |
| Nedagolla[b] | 5 | -2.85 ± 0.11 | 0.09 ± 0.09 | -2.97 ± 0.17 | -2.69 ± 0.18 | 13.0 ± 5.7 | -3.00 ± 0.11 | -0.06 ± 0.06 | -3.01 ± 0.11 | -2.72 ± 0.13 | 12.1 ± 4.0 |

Tungsten isotope ratios involving $^{183}$W were corrected for a small mass-independent analytical effect (Section A1). Uncertainties are based on the external reproducibility (2 s.d.) obtained from repeated analyses of the terrestrial standards (Table S2) or the in-run error (2 s.e.), whichever is larger. For samples with $N \geq 4$, the uncertainties represent Student-t 95% confidence intervals (95% CI), *i.e.*, $(t_{0.95,N-1} \times \text{s.d.})/\sqrt{N}$. Uncertainties induced by the correction for the analytical effect on $^{183}$W or for nucleosynthetic anomalies ($\varepsilon^{182}W_{\text{nuc. corr.}}$) (Kruijer et al., 2014a; Budde et al., 2022) as well as by the correction for cosmic ray exposure effects ($\varepsilon^{182}W_{\text{CRE-corr.}}$) (Kruijer et al., 2013) have been propagated. The samples are arranged in alphabetical order within each reservoir. *N*: number of analyses.

[a] (Spitzer et al., 2021)

[b] (Spitzer et al., 2022)

# Figures

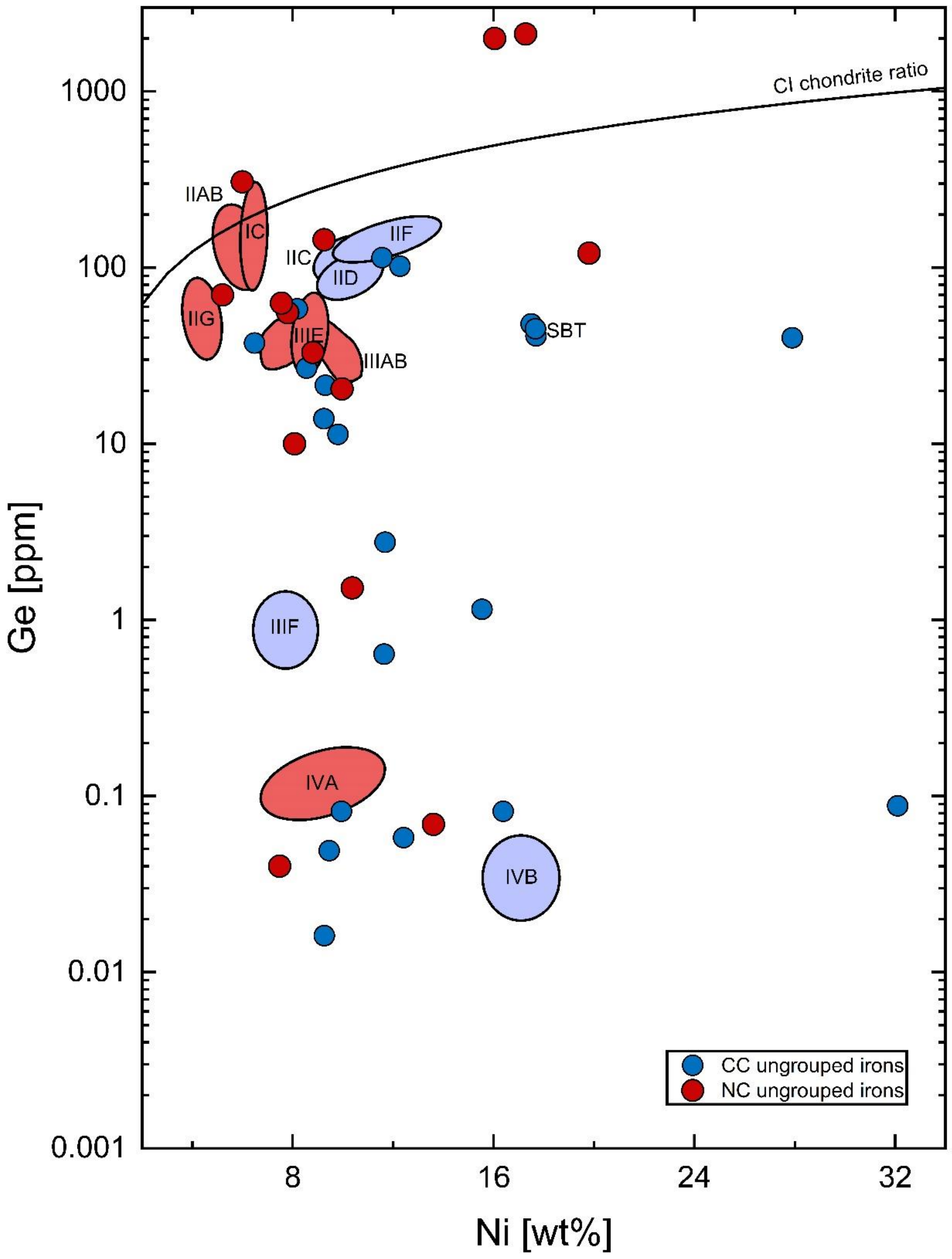

**Figure 1.** Ge vs. Ni for iron meteorites. Compositional fields of the major magmatic iron groups after Scott and Wasson (1975). The majority of the displayed ungrouped irons are from this study, but those previously identified in the literature are also shown (Table S5). CI chondrite ratio from Anders and Grevesse (1989). The chemical data for the ungrouped irons are listed in Table S1.

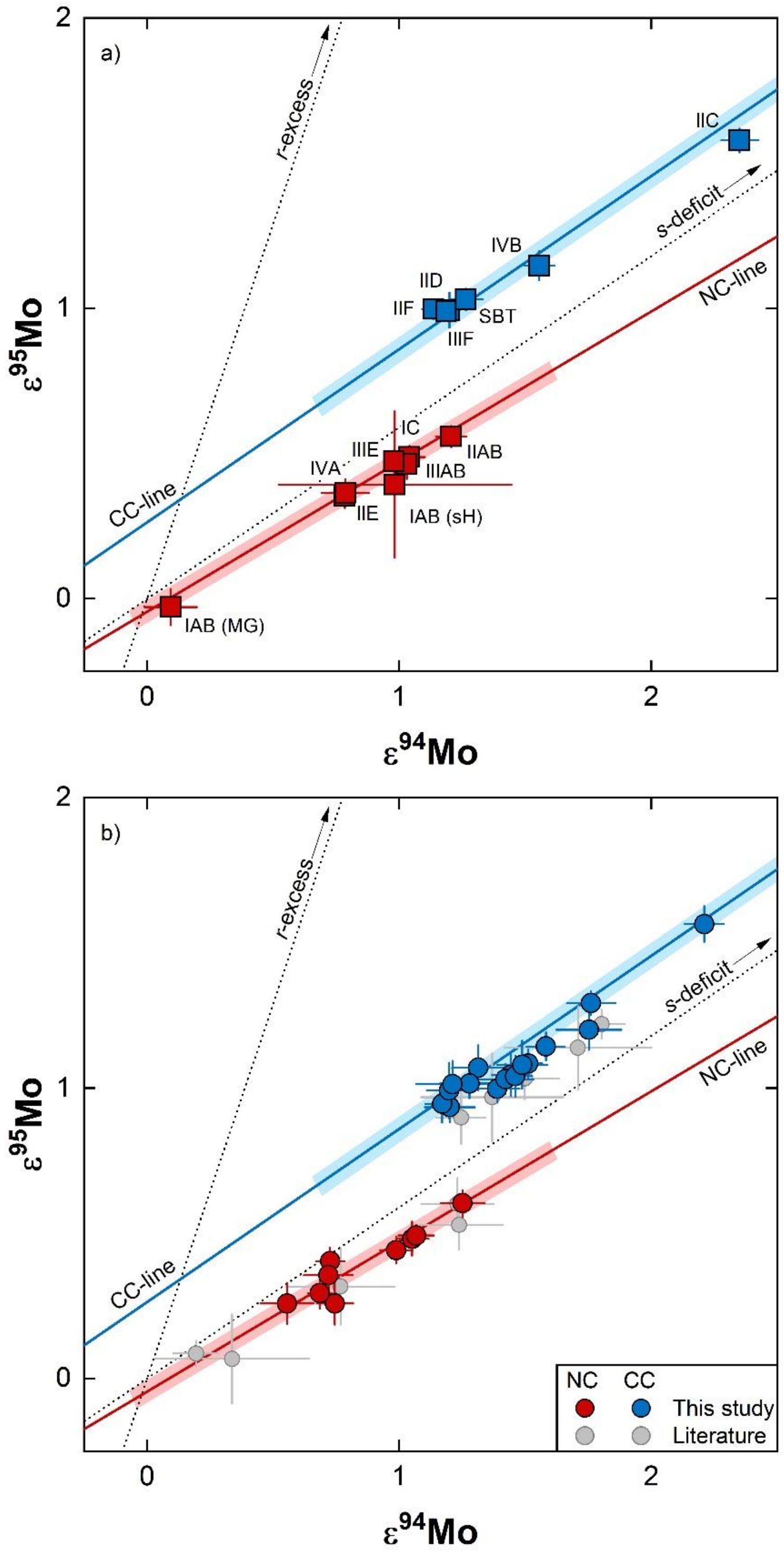

**Figure 2.** Diagram of $\varepsilon^{95}$Mo versus $\varepsilon^{94}$Mo for grouped (a) and ungrouped (b) iron meteorites. The slopes of the NC- and CC-line are from this study (see main text) and Budde et al. (2019), respectively. The shaded area represents the range of NC and CC meteorite literature data. Wiley (IIC anomalous) is not shown ($\varepsilon^{94}$Mo = 3.45). Literature data are reported in Table S5. The legend is the same for both panels.

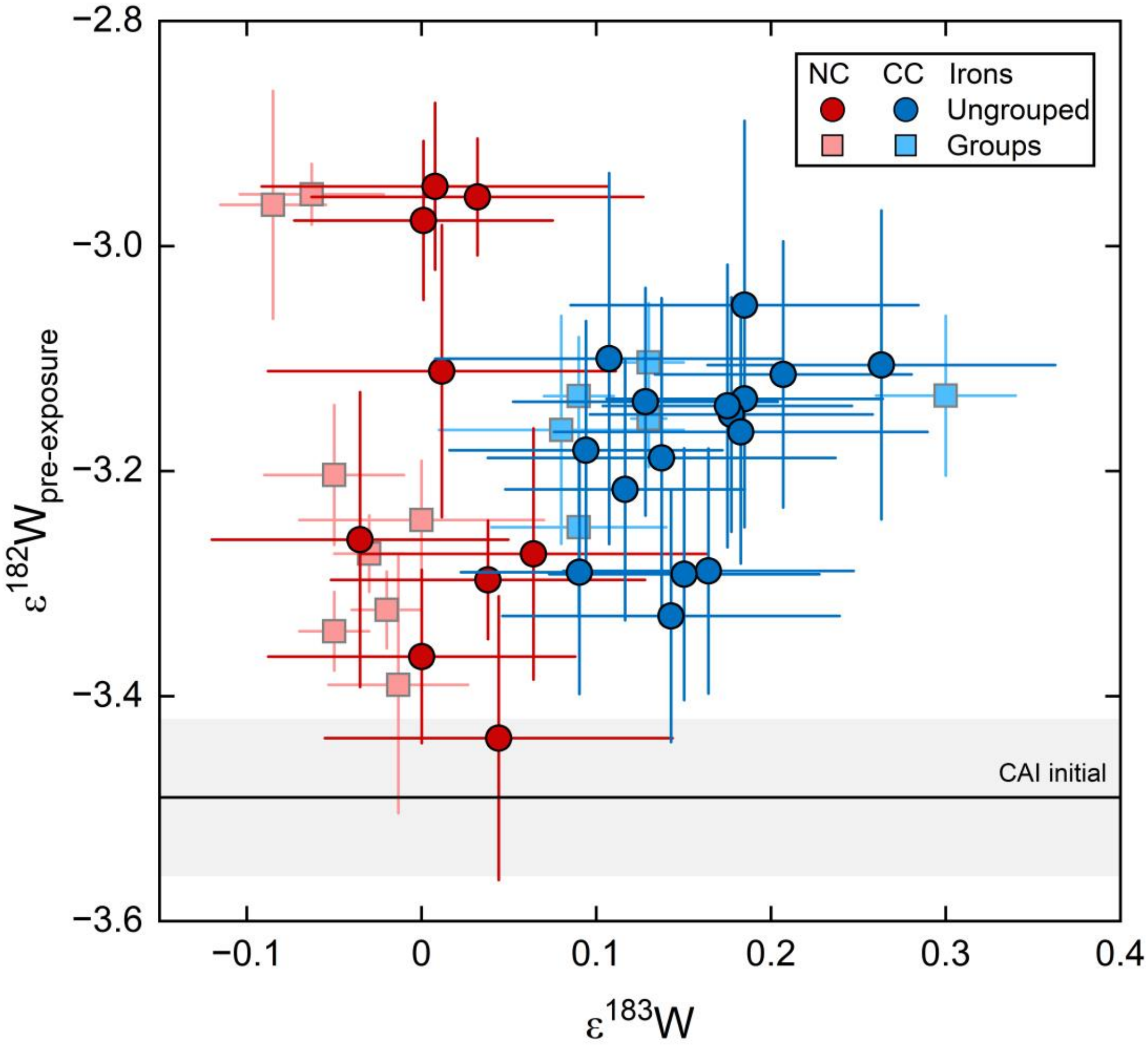

**Figure 3.** Tungsten isotope dichotomy of iron meteorites. The $\varepsilon^{182}W$ signatures were corrected for effects of nucleosynthetic heterogeneity and secondary *n*-capture (see Section 3.3 for details). CAI initial from Kruijer et al. (2014a). Literature data for iron meteorite groups are compiled in Table S4.

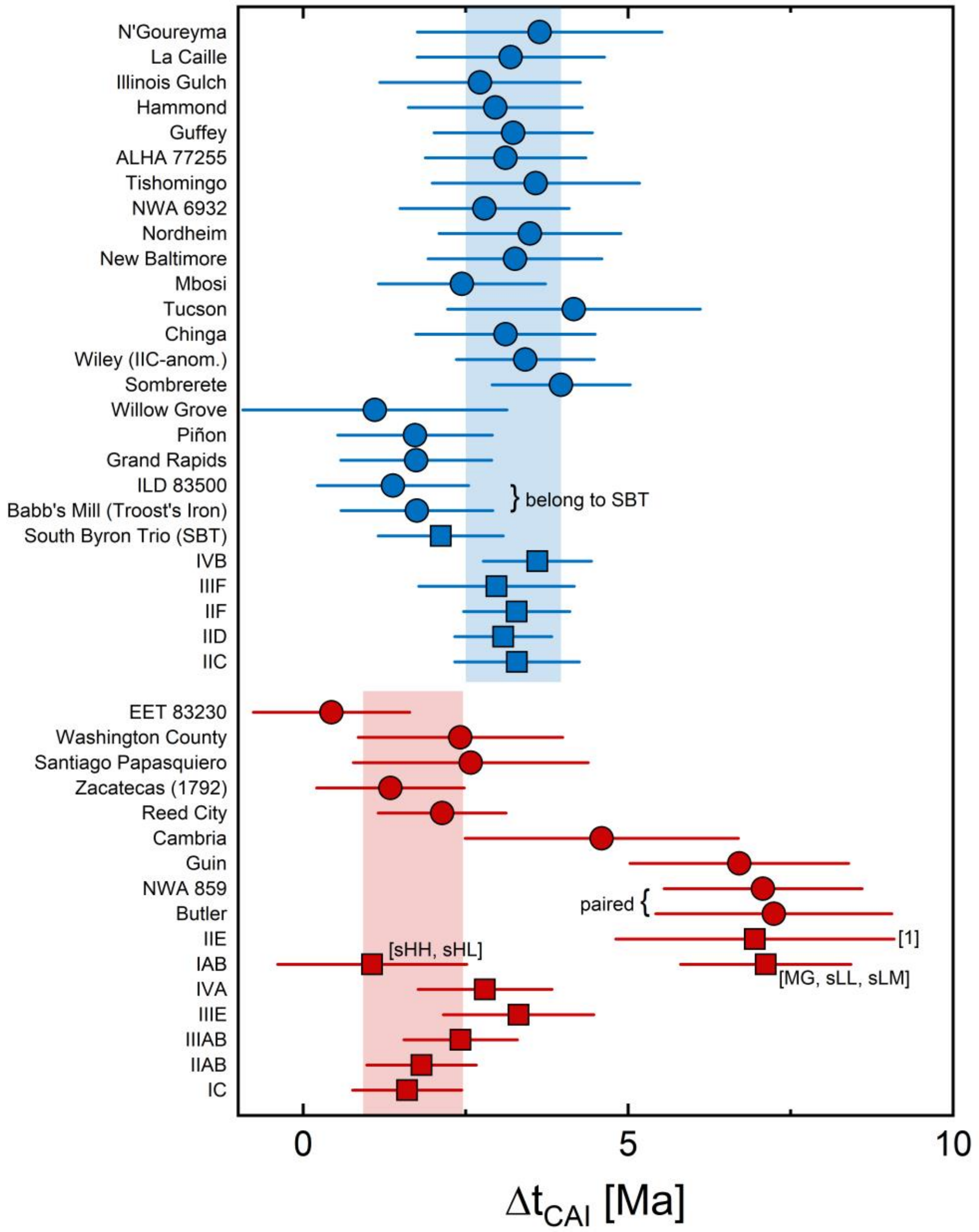

**Figure 4**. Two-stage Hf-W model ages of core formation for iron meteorite groups and ungrouped iron meteorites. Most ungrouped CC iron meteorites have core formation ages overlapping with those of the major CC groups (~3.3 Ma). Samples from the SBT together with Grand Rapids, Willow Grove, and Piñon display slightly older model ages of ~2 Ma after CAIs. On the other hand, NC iron meteorites exhibit more variable Hf-W model ages. Squares represent the major iron meteorite groups and circles ungrouped iron meteorites. Red and blue shaded areas represent the means of the volatile-rich NC (IC, IIAB) and all CC (IIC, IID, IIF, IIIF, IVB) iron meteorite groups, respectively. For the NCs, only the volatile-rich (i.e., S-rich) groups were selected because the melting temperature of metal is a function of the S content (Fei et al., 1997). Thus, for these groups, the assumption of single-stage core formation is most valid and the later extraction of silicate melt is inconsequential for estimating the relation between core formation and parent body accretion time (Kruijer et al., 2017). The pre-exposure $\varepsilon^{182}W$ signatures were corrected for effects of nucleosynthetic heterogeneity and secondary neutron capture (Section 3.3). Literature data for iron meteorite groups are compiled in Table S4.

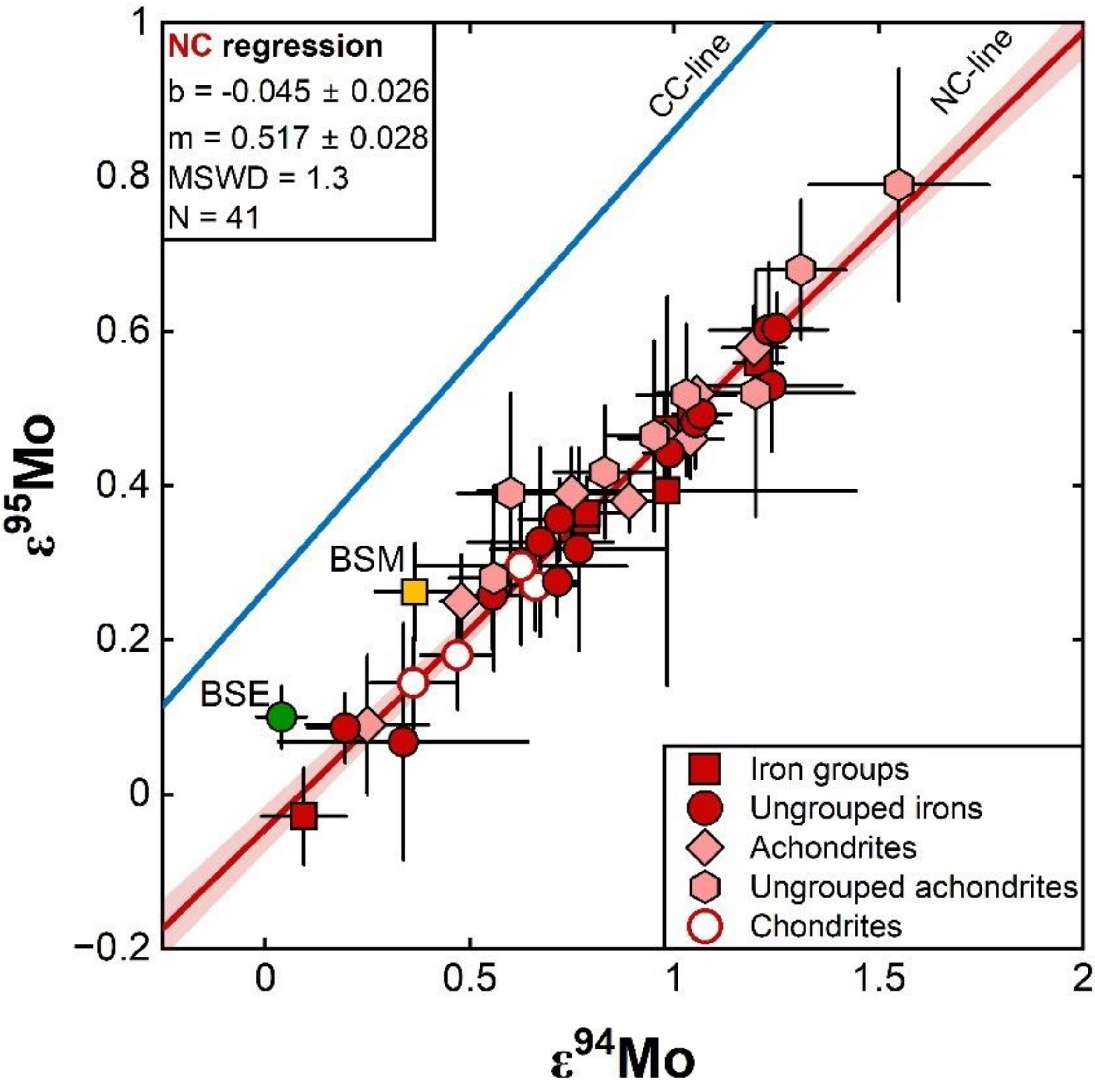

**Figure 5.** Diagram of $\varepsilon^{95}$Mo vs $\varepsilon^{94}$Mo for NC meteorites. All samples plot along a single regression line with no excess scatter and a slope slightly shallower than that expected from pure *s*-process variation indicating that NC meteorites are characterized by coupled *s*- and *r*-process variations (Spitzer et al., 2020). BSE – Bulk Silicate Earth. BSM – Bulk Silicate Mars. The Mo data are reported in Table S5. Correlated uncertainties were considered in the NC regression, but error ellipses are omitted for clarity.

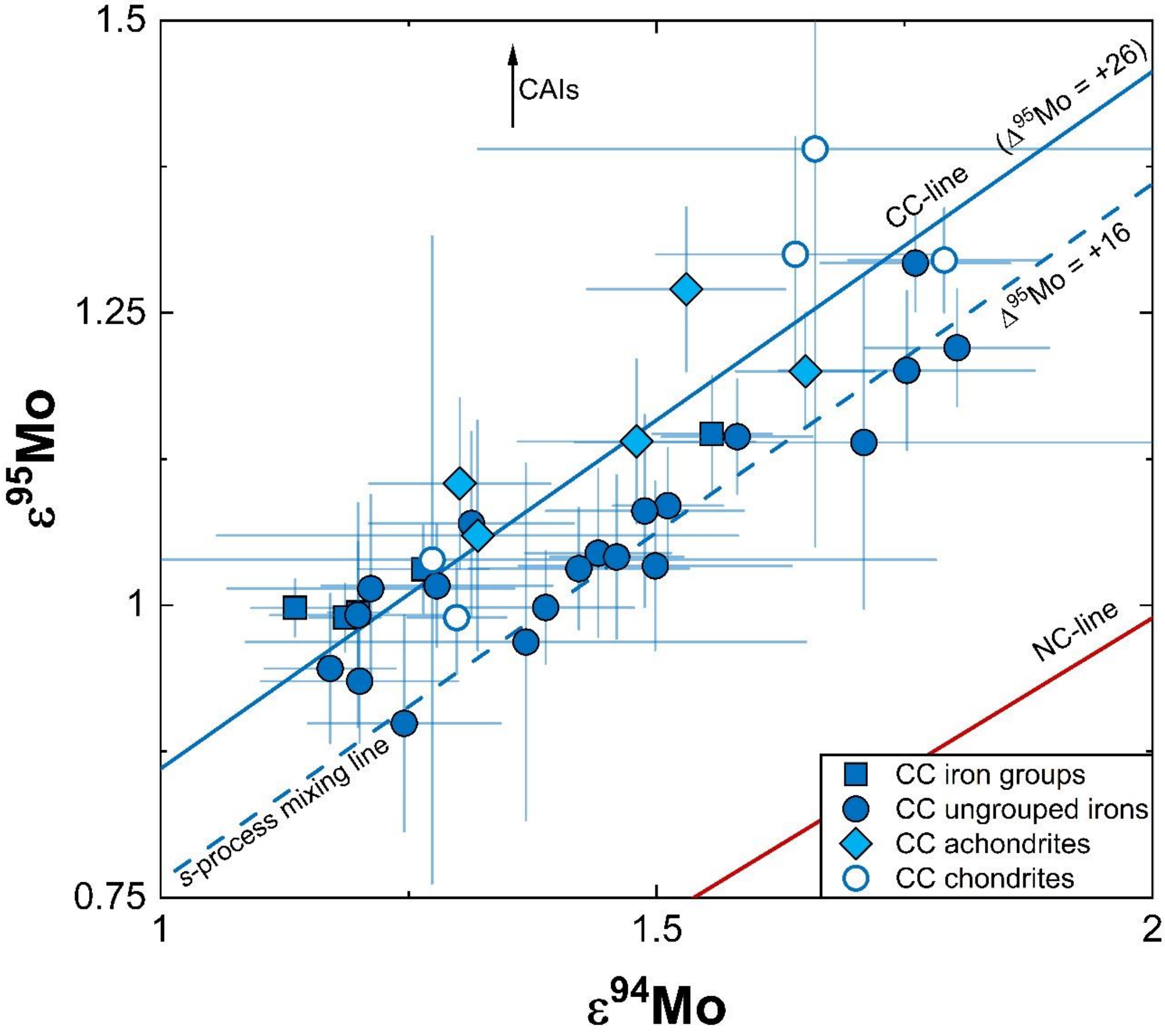

**Figure 6.** Diagram of $\varepsilon^{95}$Mo vs. $\varepsilon^{94}$Mo for CC meteorites. The $\Delta^{95}$Mo of the CC-line is from Budde et al. (2019). The blue dashed line represents an *s*-process mixing line with a $\Delta^{95}$Mo of +16 (i.e., smaller *r*-excess) demonstrating variability among CC meteorites. Several of the studied ungrouped iron meteorites plot below the CC-line defined by most CC meteorites (Budde et al., 2019). Data are compiled in Table S5.

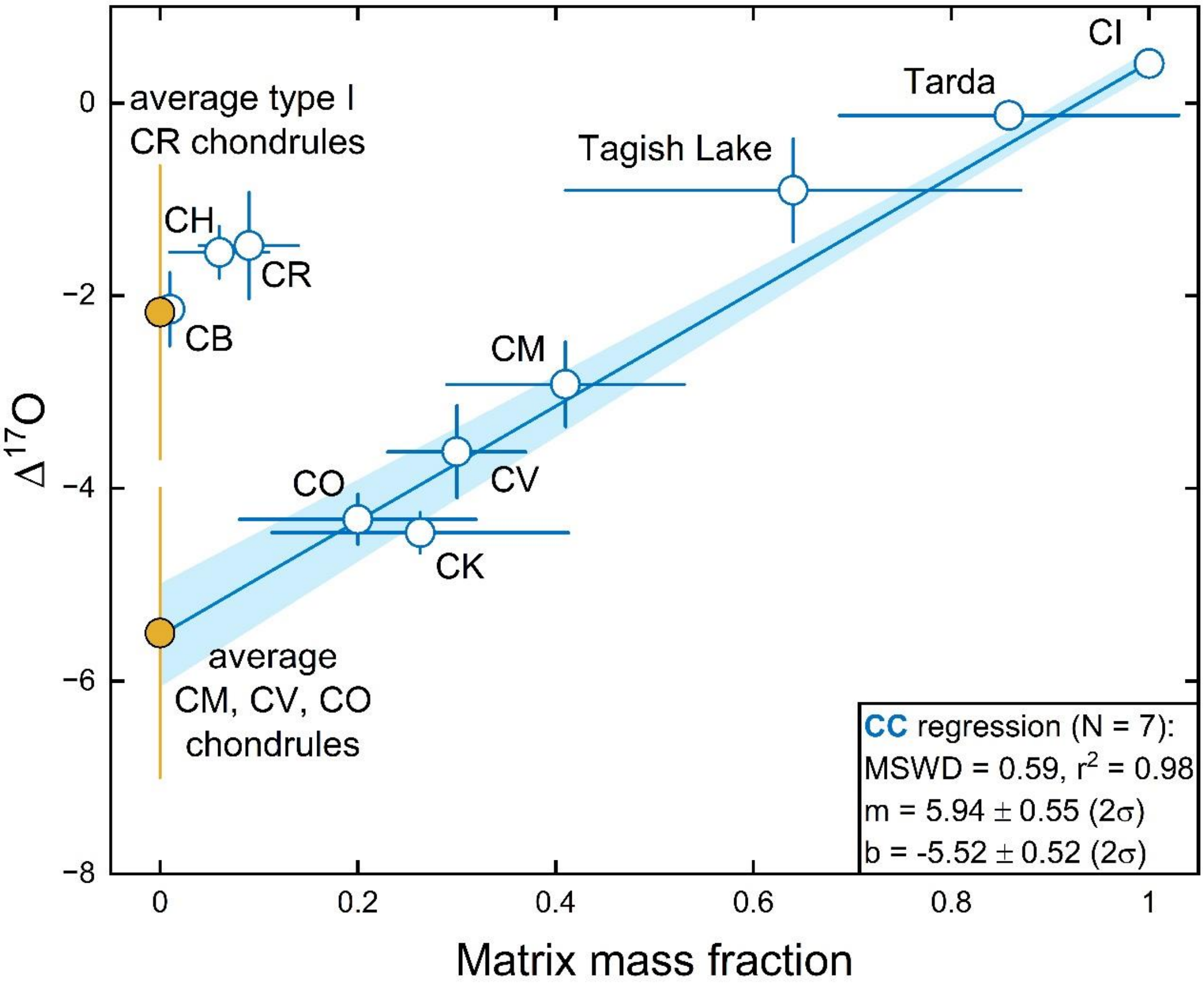

**Figure 7.** Diagram of $\Delta^{17}O$ vs. inferred matrix mass fraction for CC chondrites. These meteorites display a linear correlation consistent with a two-component mixing between a CI-like matrix and a chondrule component. The y-intercept of the regression is consistent with the measured mean O isotopic composition of CV-CM-CO chondrules. The composition of the metal-rich chondrites (CR, CH, CB) can also be explained by mixing of CI-like matrix with large type I CR chondrules, which are isotopically distinct from those of CM, CV, CO chondrules (Marrocchi et al., 2022). Data are compiled in Table S6.

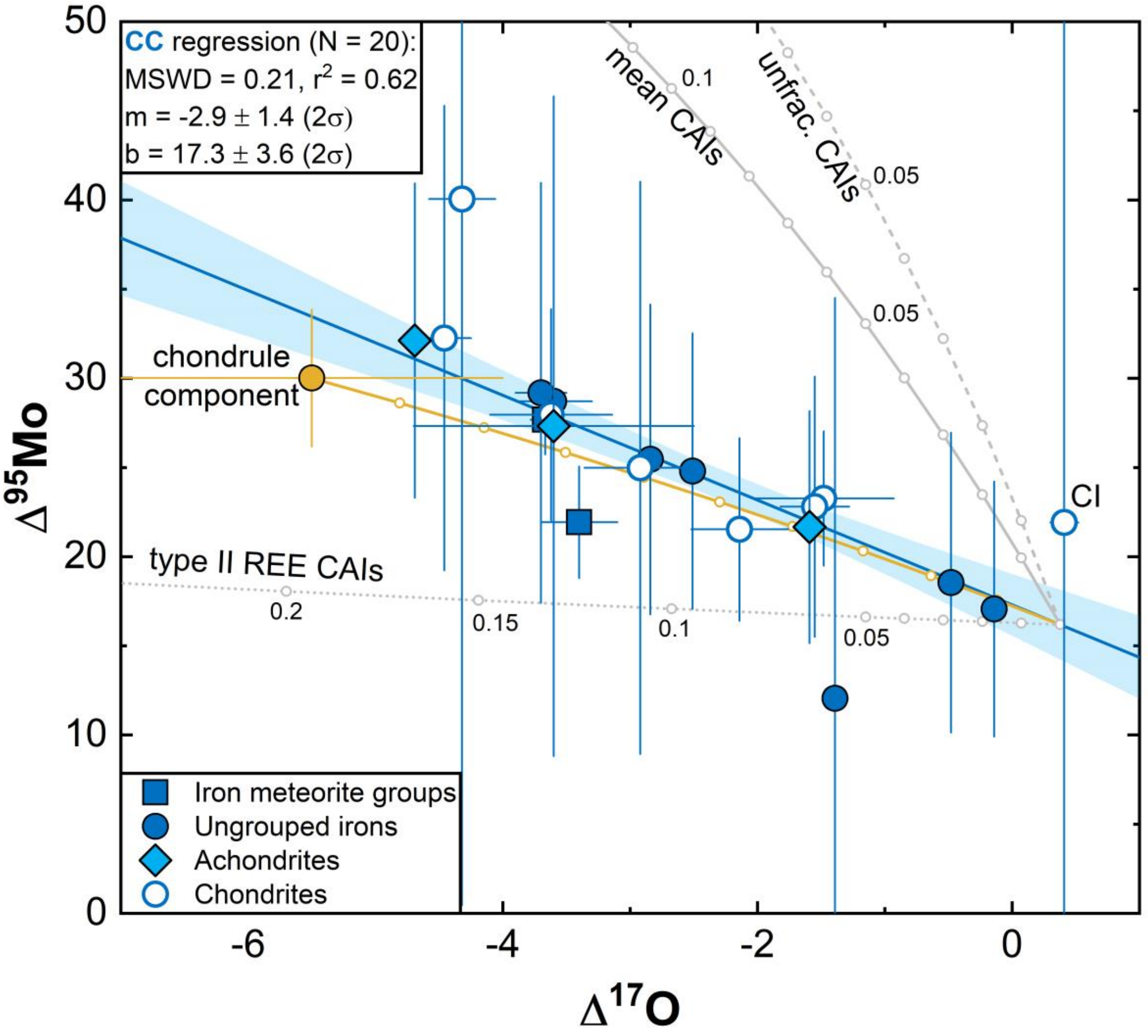

**Figure 8.** Diagram of $\Delta^{95}$Mo vs. $\Delta^{17}$O for CC meteorites, which are linearly correlated suggesting coupled mixing of CI-like matrix, chondrules, and CAIs. The chondrule component is constructed from the Mo isotopic composition of Allende chondrule separates (Budde et al., 2016a) and the O isotopic range of CV-CM-CO chondrules. The solid yellow line is a mixing line between this chondrule component and CI-like matrix with 10 wt% increments. The grey dashed, dotted, and solid mixing lines are calculated using the compositions of 'unfractionated' CAIs, group II CAIs, and average CAIs as discussed in the main text and reported in Table S7. The Mo and O data are compiled in Tables S5– 6.

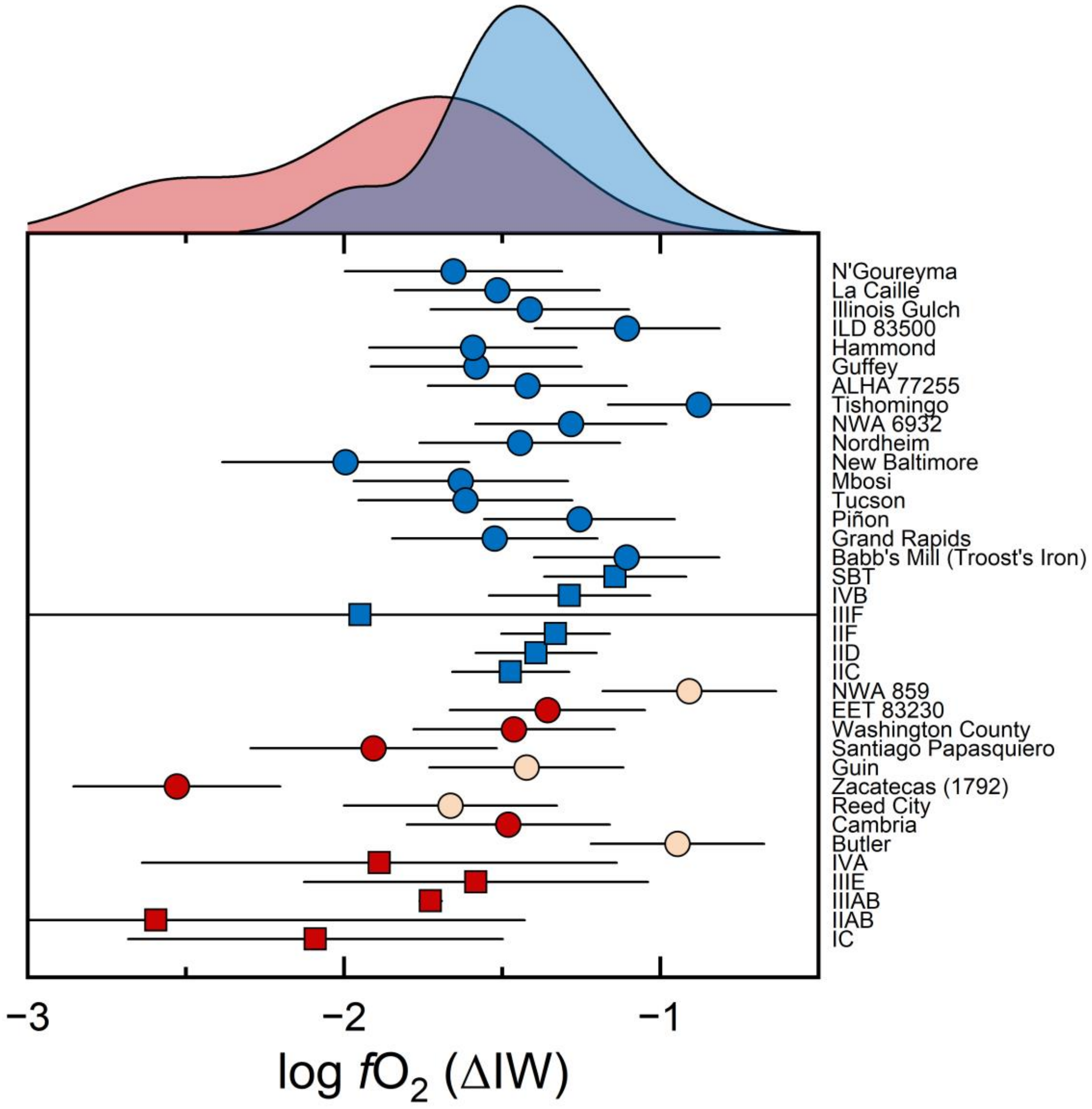

**Figure 9.** The calculated oxidation states of the iron meteorite parent bodies expressed relative to the IW buffer (see Appendix for details) display core formation conditions at IW ~-1 to -2.5. The probability density function for the CC irons shows a relatively well-defined peak at IW ~-1.4, whereas NC irons display more variable oxygen fugacities. Furthermore, the $fO_2$ values of NC and CC iron meteorites overlap to some extent. Nevertheless, a two-tailed unequal variances t-test reveals that the difference between their means is statistically significant at the 95% confidence limit indicating that parent bodies of CC irons on average formed under more oxidizing conditions than their NC counterparts. The light red circles represent ungrouped iron meteorites, which likely are of non-magmatic origin (see main text for details) and are therefore not included in the calculations. The error bars reflect the 2 s.d. of the two estimates obtained from the Fe/Ni and Fe/Co ratios, respectively (Table S8).

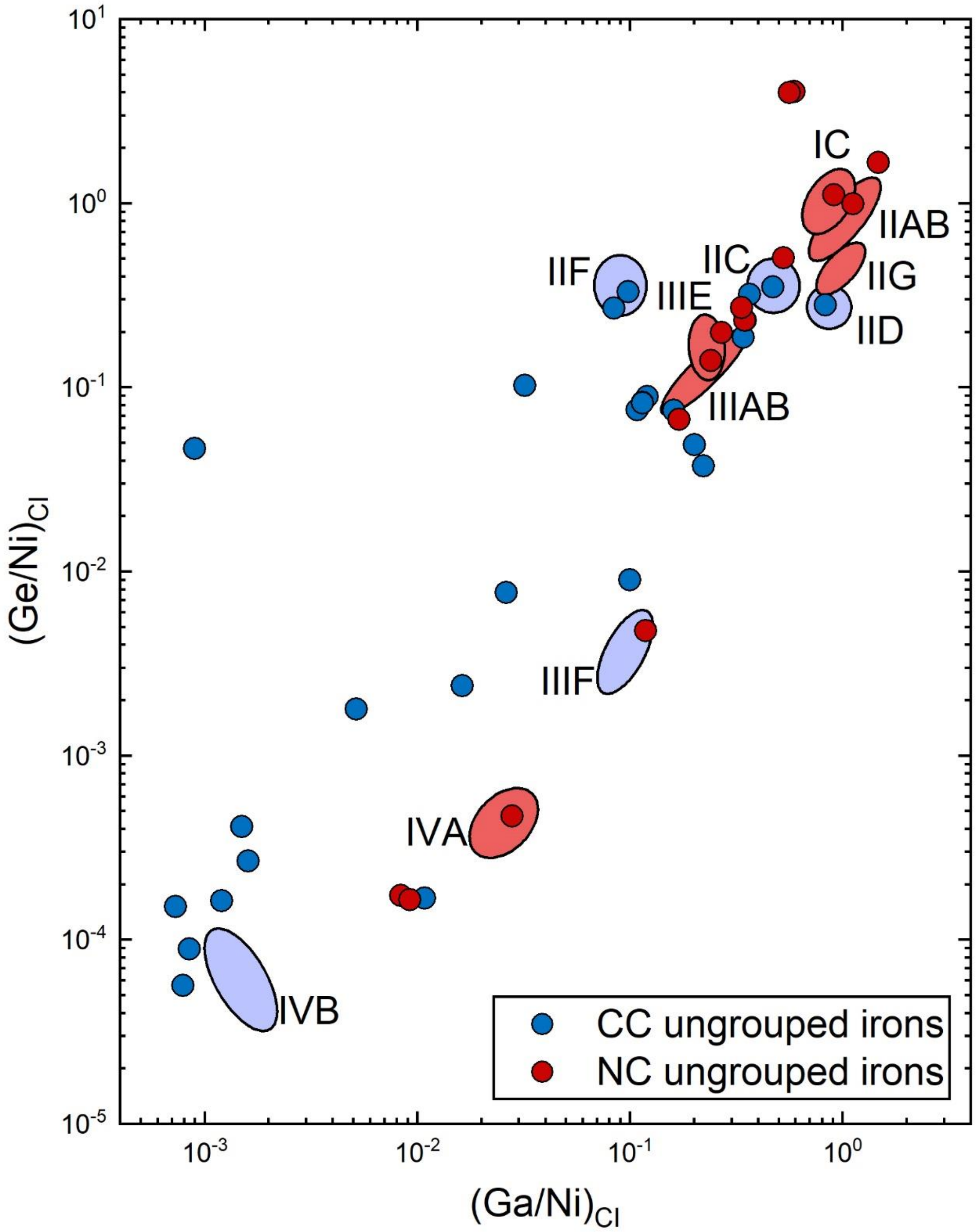

**Figure 10.** CI-normalized Ge/Ni vs. Ga/Ni ratios for iron meteorites illustrating the variable degrees of volatile depletion among different groups of iron meteorites. Ungrouped iron meteorites mirror the trend displayed by the major magmatic iron meteorite groups. Combined, they reveal a higher abundance of volatile-depleted samples among CC than NC bodies. Data of the ungrouped irons are listed in Table S1.